\documentclass[aps,superscriptaddress,preprintnumbers,floatfix,twocolumn,amsmath,amssymb]{revtex4-2}
\usepackage{graphicx}
\graphicspath{{Pictures/}}
\usepackage{hyperref}
\hypersetup{pdfnewwindow=true, colorlinks=true, linkcolor=magenta, anchorcolor=blue, citecolor=red, filecolor=blue, menucolor=blue, urlcolor=blue}
\usepackage{cleveref}
\usepackage{color}
\usepackage{float}
\usepackage{soul}
\usepackage{ulem}
\usepackage{braket}
\usepackage{MnSymbol}
\usepackage{bm} 

\makeatletter

\newcommand{\vk}{\ensuremath{\mathbf{k}}}
\newcommand{\vq}{\ensuremath{\mathbf{q}}}
\newcommand{\vp}{\ensuremath{\mathbf{p}}}
\newcommand{\dr}{\ensuremath{\mathbf{r}}}
\newcommand{\bS}{\ensuremath{\mathbf{S}}}
\newcommand{\vs}{\ensuremath{\mathbf{S}}}

\begin{document}

\title{Electromagnetic signatures of chiral quantum spin liquid}

\author{Saikat Banerjee}
\affiliation{Theoretical Division, T-4, Los Alamos National Laboratory, Los Alamos, New Mexico 87545, USA}
\author{Wei Zhu}
\affiliation{School of Science, Westlake University, No. 600 Dunyu Road, Hangzhou 310030, China}
\affiliation{Key Laboratory for Quantum Materials of Zhejiang Province, Westlake University, Hangzhou 310024, China}
\author{Shi-Zeng Lin} 
\affiliation{Theoretical Division, T-4, and CNLS, Los Alamos National Laboratory, Los Alamos, New Mexico 87545, USA}
\affiliation{Center for Integrated Nanotechnologies (CINT), Los Alamos National Laboratory, Los Alamos, New Mexico 87545, USA}

\date{\today}
\begin{abstract}
Quantum spin liquid (QSL) has become an exciting topic in interacting spin systems that do not order magnetically down to the lowest experimentally accessible temperature; however, conclusive experimental evidence remains lacking. Motivated by the recent surge of theoretical and experimental interest in a half-filled Hubbard model on the triangular lattice, where chiral QSL can be stabilized, we investigate the electromagnetic signature of the chiral QSL to aid experimental detection. We systematically studied the electrical charge and orbital electrical current associated with a spinon excitation in the chiral QSL based on parton mean-field theory and unbiased density-matrix renormalization group calculations. We then calculated both longitudinal and transverse optical conductivities below the Mott gap. We also conduct quantum field theory analysis to unravel the connection between spinon excitation and emergent and physical gauge fields. Our results show that the chiral QSL phase has a clear electromagnetic response even in a Mott insulator regime, which can facilitate the experimental detection of this long-sought-after phase.
\end{abstract}

\maketitle

\section{Introduction \label{sec:sec_I}}

Quantum spin liquid (QSL) states are interacting quantum spin systems that do not order magnetically down to zero temperature. This absence of magnetic order leads to a quantum-disordered ground state with characteristic long-range quantum entanglement, fractionalized excitations, and its associated emergent gauge fields. Consequently, it has been challenging to understand and characterize QSL since its inception~\cite{Anderson1973}. The experimental detection of QSL states becomes even more difficult due to the lack of a conventional order parameter~\cite{Broholm2020}. However, recent developments in both the theoretical and experimental fronts have led to a continuous surge of interest in analyzing and detecting this illusive state of matter~\cite{RevModPhys.89.025003}. Examples range from the discovery of various iridates/ruthenates compounds as candidate materials to realize proximate Kitaev physics~\cite{Kitaev2006,Savary_2017} to the observation of topological spin liquids in the Rydberg atom quantum simulator~\cite{Semeghini2021} and quantum processor \cite{Satzinger_2021}. The appearance of QSL requires suppressing the magnetic orders, and therefore frustrated magnets are the playground for hunting for QSL. In this regard, the triangular lattice Hubbard model (TLHM) has always remained a centerpiece of attention.

\begin{figure}[b]
\centering
\includegraphics[width=0.79\linewidth]{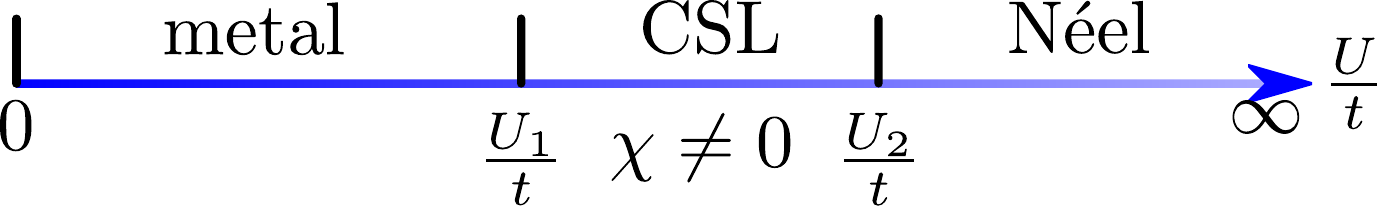} 
\caption{A schematic phase diagram for the triangular lattice Hubbard model at half-filling with a metallic phase at small $U$, followed by a putative cQSL phase with non-vanishing chiral order parameter $\chi = \braket{\mathbf{S}_i \cdot ( \mathbf{S}_j \times \mathbf{S}_k}$ at an intermediate coupling regime $U_1 \sim 9t$, and $U_2 \sim 11t$~\cite{PhysRevB.106.094420,PhysRevX.10.021042}, and a magnetic ordered N\'eel state at strong coupling. Note that $\chi = 0$ in the other two phases.} \label{fig:Fig1}
\end{figure}

In the large $U$ limit of the TLHM at half-filling, the effective low energy Hamiltonian is an antiferromagnetic Heisenberg model which stabilizes the conventional $120^\circ$ (N\'eel) order \cite{Bernu1992,Sorella1999,WJHu2015}. However, it is widely believed that the ground state of the TLHM drifts toward a QSL state when the correlations become weaker but remain above the Mott transition~\cite{Sorella2012}. Recently, various density-matrix renormalization group studies (DMRG)~\cite{PhysRevX.10.021042,PhysRevLett.127.087201,PhysRevB.105.205110,PhysRevB.106.094420}, and matrix product state (MPS)~\cite{PhysRevB.106.094417} analyses on TLHMs have predicted the evidence for a Kalmeyer-Laughlin type chiral quantum spin liquid (cQSL) phases~\cite{PhysRevLett.59.2095}, see Fig.~\ref{fig:Fig1} for a schematic phase diagram.

The TLHM can be realized in certain materials. Previous experimental work has shown characteristic evidence for a QSL phase in certain organic Mott insulators~\cite{PhysRevLett.91.107001,PhysRevB.77.104413,Miksch,Andrej2022}. Although, the controversy over the gapped~\cite{Miksch} or gapless~\cite{Li2015} nature of the underlying excitations still remains. In another triangular lattice material, $\mathrm{YbMgGaO_4}$, the gapless character is well supported by the nuclear magnetic resonance~\cite{PhysRevB.102.045149} and muon spin rotation~\cite{PhysRevB.100.241116} experiments, as well as evidence of a spinon Fermi surface revealed by neutron scattering studies~\cite{Shen2016}. Therefore, it is necessary to look for some \textit{smoking-gun} signatures that can decipher the true nature of the QSL phase.

Motivated by the identification of the cQSL phase in TLHM and its potential relevance in several compounds, here, we systematically analyze its electromagnetic responses. Despite being a Mott insulator, there is a remnant electromagnetic response due to the virtual hopping of electrons~\cite{PhysRevB.73.155115,PhysRevB.78.024402}. Assuming a cQSL phase, which spontaneously breaks time-reversal symmetry (TRS), we analyze the corresponding effective spin model~\cite{PhysRevLett.127.087201} within the parton mean-field spinon description and obtain the associated orbital magnetization and electrical polarization. We also performed unbiased DMRG calculations on the half-filled TLHM at an intermediate coupling $U$ ($U_1 < U < U_2$). Our numerical analysis further supports the mean-field results for the electromagnetic responses. To have a universal picture, we additionally employ the quantum field theory description to elucidate explicitly the relationship among the emergent and the physical gauge fields and low energy spinon excitations in the cQSL. 

To relate our theoretical framework to experiments, we compute the transverse optical conductivity (within the spinon description), which is associated with the magneto-optical Faraday rotation (MOFE)
\begin{equation}\label{eq.1}
\Theta_{\rm{F}} = \frac{l}{nc} \sigma'_{xy} (\Omega),
\end{equation}
where $l$ is the thickness in the direction of light propagation with frequency $\Omega$, $n$ is the index of refraction, and $\sigma'_{xy} (\Omega)$ is the real part of the optical conductivity in 3D. Our electromagnetic response functions, including the orbital magnetization profile and the structure of $\Theta_{\rm{F}}$, provide a clear experimental signature of the cQSL. For completeness, we also analyze the behavior of the dynamic spin-structure factor and lay out the possible connection with the relevant experiments.

The rest of this paper is organized as follows: in Sec.~\ref{sec:sec_II}, we provide the spinon description of the cQSL in TLHM. In Sec.~\ref{sec:sec_III} and Sec.~\ref{sec:sec_III.I}, we provide details of the derivation for electrical polarization and orbital magnetization. The DMRG calculations supporting our mean-field calculations are given in Sec.~\ref{sec:sec_III.WZ}. Sec. \ref{sec:sec_III.QFT} provides a picture based on quantum field theory. In Sec.~\ref{sec:sec_III.II}, and Sec.~\ref{sec:sec_III.III}, we compute the dynamic spin structure factor and transverse optical conductivity with the electrical polarization, respectively. Finally, we discuss the implications of our results and conclude in Sec.~\ref{sec:sec_IV}.

\section{Model \label{sec:sec_II}}

We start with the TLHM at half-filling with the corresponding Hamiltonian written as 
\begin{equation}\label{eq.2}
\mathcal{H}_0 = -t \sum_{\langle ij \rangle, \sigma} c^{\dag}_{i\sigma} c_{j \sigma} + U \sum_{i} n_{i \uparrow} n_{i\downarrow},
\end{equation}
where $c^{\dag}_{i\sigma}$ creates an electron at site $i$ with spin $\sigma$, and $U$ is the strength of the onsite Coulomb repulsion. In the strong coupling limit ($U \gg t$), the charge degrees of freedom are gapped out, and the relevant microscopic model can be analyzed in terms of an effective spin model. Within a second-order perturbation expansion in $t/U$, the corresponding spin Hamiltonian reads $\mathcal{H}_{\rm{eff}} = J^{(2)} \sum_{\langle ij \rangle} \bS_{i} \cdot \bS_j$,
where $J^{(2)} = 4t^2/U$ is the antiferromagnetic Heisenberg coupling. However, in the intermediate coupling regime, i.e., $U \gtrsim t$, the above second-order perturbation does not completely capture the low-energy dynamics, and we need to include higher-order spin corrections. Such a procedure leads to further neighbor spin exchange terms, including ring exchange-like interactions~\cite{PhysRevB.72.115114}. Therefore, although a N\'eel order is preferred at larger $U$, incorporating subleading order correction modifies the overall magnetic order at an intermediate $U$. Previous theoretical works~\cite{PhysRevB.73.155115,PhysRevLett.103.036401,PhysRevLett.100.136402,PhysRevLett.105.267204} have reported the existence of two critical coupling strengths $U_1 \sim 9 t$, and $U_2 \sim 11t$. The current consensus is that TLHM hosts a putative QSL phase in the intermediate regime between $U_1$ and $U_2$, eventually becoming a  N\'eel ordered state at a larger $U>U_2$.

Motivated by these previous studies and recent developments in the DMRG results~\cite{PhysRevX.10.021042,PhysRevB.106.094420}, we adopt a \textit{phenomenological} chiral spin liquid model to describe its concomitant features. The effective Hamiltonian, which hosts cQSL as a ground state, is written as 
\begin{equation}\label{eq.3}
\mathcal{H}_{\rm{csl}} = \tilde{J} \sum_{\langle ij \rangle} \bS_{i} \cdot\bS_{j} + \tilde{J}_{\chi} \sum_{\llangle ijk \rrangle} \bS_{i} \cdot (\bS_{j} \times \bS_{k}),
\end{equation}
where the associated exchange couplings are written in terms of the parameters of the original low-energy spin model. Here $\langle ij \rangle$ denotes the nearest sites and $\llangle ijk \rrangle$ denotes three sites in a unit triangle. It was argued in Refs.~\cite{PhysRevLett.127.087201, PhysRevB.73.155115} that the four-spin ring exchange term (see SM~\cite{supp} for details) is responsible for the appearance of the chiral term in Eq.~\eqref{eq.3}.
	
Here, we focus on the model as in Eq.~\eqref{eq.3} and analyze it within a mean-field description. We utilize the standard parton decomposition of the spins as $\bS_i = \tfrac{1}{2} f^{\dag}_{i\alpha} \bm{\sigma}_{\alpha \beta} f_{i \beta}$, where $f^{\dag}_{i\alpha}$ creates a neutral spinon excitation with spin $\alpha$ at site $i$, and $\bm{\sigma}$ denotes the vector of Pauli matrices (the repeated indices are assumed to be summed over). This fractionalization leads to an enlargement of the Hilbert space. Therefore, one needs to implement a local constraint (${f^{\dag}_{i\alpha} f_{i\alpha}} = 1$) to project to the physical Hilbert space.  Plugging this back into Eq.~\eqref{eq.3} and assuming a nonzero mean-field decomposition as $m_{ij} = \braket{f^{\dag}_{i\alpha} f_{j\alpha}}$, we obtain a noninteracting spinon Hamiltonian as (see supplementary material (SM)~\cite{supp} for details)
\begin{equation}\label{eq.4}
\mathcal{H} = 
-\frac{\tilde{J}}{2} \sum_{\langle ij \rangle} m_{ji} f^{\dag}_{i\alpha} f_{j\alpha} 
+ \frac{3i\tilde{J}_{\chi}}{16} \sum'_{\langle ij \rangle} m_{ik} m_{kj} f^{\dag}_{j\alpha} f_{i\alpha} + \rm{h.c.},
\end{equation}
where the primed summation corresponds to all the permutations between the three neighboring sites $i,j,k$. Here, we adopted a mean-field decomposition only in the particle-hole channel, although a more general decomposition with both particle-particle and particle-hole channel may provide a \textit{qualitatively} better description of the emergent spinon spectrum~\cite{Mezio_2011,PhysRevB.79.014424}. 

\begin{figure}[t]
\centering
\includegraphics[width=1\linewidth]{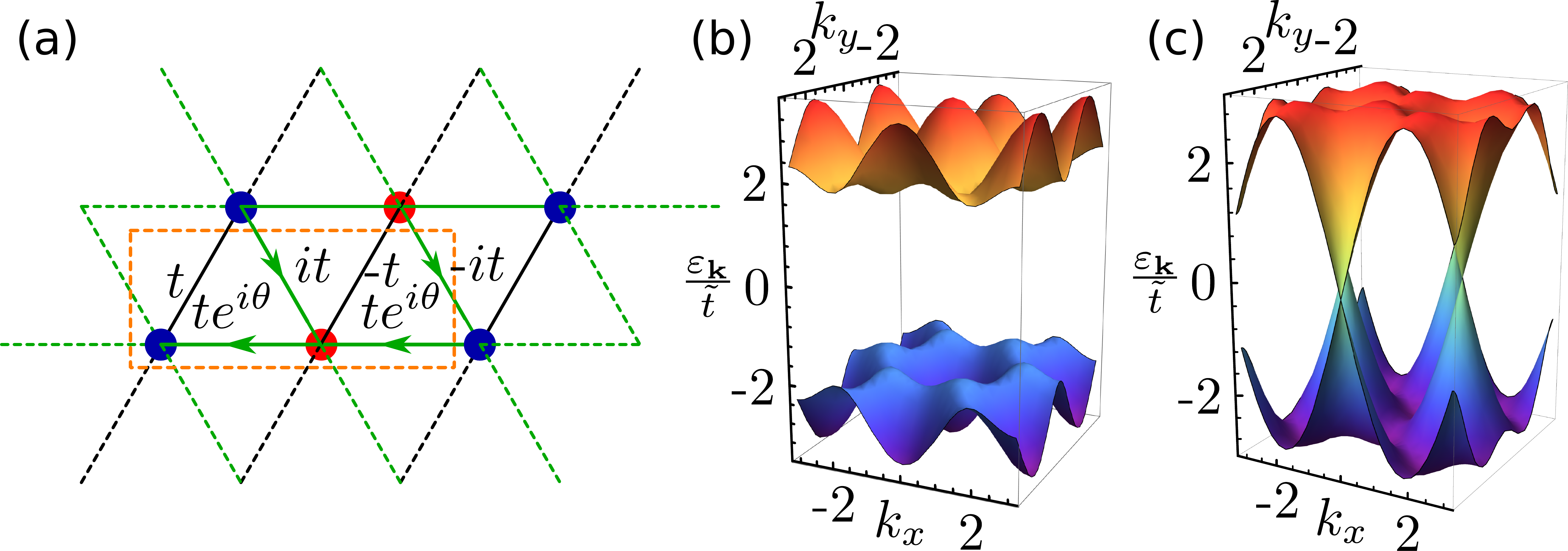} 
\caption{(a) Phenomenological spinon model on a triangular lattice with bond-dependent hoppings and a two-sublattice unit cell illustrated within the orange-dashed box. The hopping phases allow $\pi$-flux within each rhombus-shaped bipartite plaquette (see the main text for more discussion). For $\theta = 0$, both the \textit{up} and a \textit{down} triangle forming the rhombus acquire uniform $\pi/2$ fluxes, where the flux configuration is staggered for any nonzero $\theta$. The spinon spectrum for the uniform (gapped, $\theta = 0$) and the staggered (gapless, $\theta = \pi/2$) flux configuration are shown in panels (b) and (c), respectively.}\label{fig:Fig2}
\end{figure}

Assuming the translational invariance, we simplify the mean-field order parameter $m_{ij} = m_0 e^{i \phi_{ij}}$, where $m_0$ is the amplitude, and $\phi_{ij}$'s are bond-dependent phases. Subsequently, we capture the physics of the Hamiltonian in Eq.~\eqref{eq.4} with a simplified model as
\begin{equation}\label{eq.5}
\mathcal{H} = - \tilde{t} \sum_{\langle ij \rangle} e^{i\psi_{ij}} f^{\dag}_{i\alpha} f_{j\alpha} + \rm{h.c.}.
\end{equation}
Focusing on a three-site cluster, the hopping amplitude $\tilde{t}$, and the phases $\psi_{ij}$'s are related to the parameters in Eq.~\eqref{eq.4} as 
\begin{subequations}
\begin{align}
\label{eq.6.1}
\tilde{t} \cos \psi_{ij} & = \frac{\tilde{J} m_0}{2} \cos \phi_{ji} + \frac{3 \tilde{J}_{\chi} m_0^2}{16} \sin \left( \phi_{ik} + \phi_{kj} \right), \\
\label{eq.6.2} 
\tilde{t} \sin \psi_{ij} & = \frac{\tilde{J} m_0}{2} \sin \phi_{ji} + \frac{3 \tilde{J}_{\chi} m_0^2}{16} \cos \left( \phi_{ik} + \phi_{kj} \right),
\end{align}	
\end{subequations}
However, the phases $\psi_{ij}$'s and the hopping $\tilde{t}$ remain undetermined. To further progress, we utilize Lieb's theorem~\cite{PhysRevLett.73.2158}, which states that a fermion hopping on a bipartite lattice realizes its ground state with $\pi$-flux per bipartite plaquettes. Since the triangular lattice is monopartite, we consider a decorated lattice comprised of doubled unit cells [see Fig.~\ref{fig:Fig2}(a)] with the hopping amplitudes between different neighboring sites such that the total flux within the rhombus-shaped unit cell is $\pi$. In such a construction, we can do further simplification and solve Eq.~\eqref{eq.6.1}, and Eq.~\eqref{eq.6.2} to show that~\cite{supp}
\begin{equation}\label{eq.7}
\tilde{t} = \frac{\tilde{J} m_0}{2} + \frac{3 \tilde{J}_{\chi}m_0^2}{16}, \quad \psi_{ij} = - \phi_{ij}
\end{equation}
with the constraint, the total flux within a triangle is $\pi/2$. Note that $m_0$ still remains undetermined. A particular choice of $\psi_{ij}$ is shown in Fig.~\ref{fig:Fig2}(a) to realize the staggered flux configurations between the up and the down triangles, where $\theta=0$ corresponds to $\pi/2$ flux within a triangle. TRS is preserved for $\theta = \pi/2$. 

Diagonalizing the Hamiltonian in Eq.~\eqref{eq.4} obtains the corresponding spinon band structure. The uniform flux phase ($\theta = 0$) leads to a gapped spinon spectrum, as shown in Fig.~\ref{fig:Fig2}(b). Note that the spectrum becomes gapless for the staggered flux configuration with $\theta = \pi/2$, and remains gapped for any other choice of $\theta$. The spinon spectrum is doubly degenerate for the spin-up and spin-down components. The gapped bands acquire a nonzero Chern number in the uniform flux configuration. Using the link variable formulation~\cite{Fukui2005}, we obtain the total Chern number distribution for the bands as $\mathcal{C} = \{2, -2\}$ in the cQSL phase. Therefore, it is expected to host \textit{chiral} spinon edge modes and exhibit quantized Hall thermal conductivity at low temperatures~\cite{Nakai_2016}.

\begin{figure*}[t]
\centering
\includegraphics[width=1\linewidth]{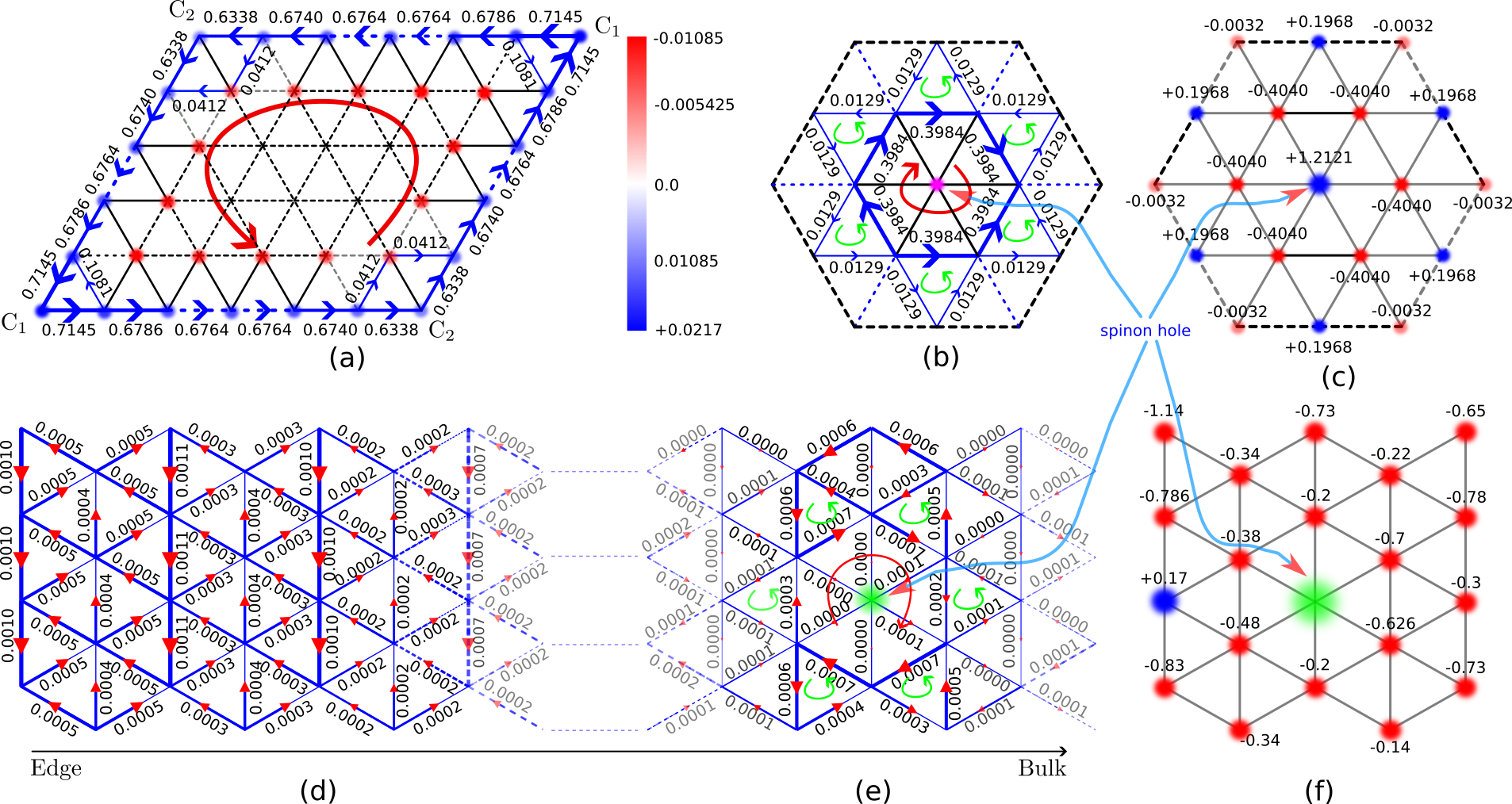} 
\caption{(a) An illustration of the localized loop current and charge distributions around the edge of a finite system of linear size $L = 30$ (in open boundary condition) within the mean-field spinon description of the spin model in Eq.~\eqref{eq.3}. For illustrative purposes, we do not show the explicit distribution of the loop currents within the bulk. Note that the loop current and charge fluctuation quickly vanish after a few lattice spacings inside the bulk. The loop currents (b) and charge distribution (c) around the localized spinon hole site were obtained in periodic boundary conditions with the same system size. The red and blue colors signify the opposite signs of charge redistribution. The numbers are presented in the unit of $2\mathcal{I}_0$ and $2\rho_0$, respectively (see the main text). (d) and (e), Plots of the local electric currents on the triangular Hubbard model for (d) without and (e) with a spinon hole located at the position labeled by the green color obtained by the DMRG calculations. The loop current emerges around the local magnetic field. The red arrows represent the direction of the loop current. The numbers around the bonds label the absolute value of the current in the unit of $et/\hbar$. (f) Plot of the charge redistribution around the local spinon hole on a finite-size system (illustrated by the green region) obtained by the DMRG calculations. For clarity, the numbers are in the unit of $10^{-4}e$.}\label{fig:Fig3}
\end{figure*} 

\section{Analysis and Results \label{sec:sec_III}}

Now we discuss the main results of this paper by focusing on the electromagnetic signatures in the cQSL phase. Despite a charge-neutral Mott insulator, the virtual hopping of electrons leads to a nonvanishing expectation value of the charge fluctuations and circulating loop currents in the cQSL phase~\cite{PhysRevB.78.024402}. In fact, such features are expected in spin liquid systems~\cite{PhysRevB.87.245106,PhysRevLett.125.227202,Banerjee2022}. The relevant operators for the charge fluctuations and loop currents in the TLHM read~\cite{PhysRevB.78.024402}
\begin{subequations} 
\begin{align}
\label{eq.8.1}
\delta \hat{\bm{\rho}}_{i,jk}			&	=	e \frac{8t^3}{U^3}  \left( \mathbf{S}_i \cdot \mathbf{S}_j + \mathbf{S}_i \cdot \mathbf{S}_k - 2 \mathbf{S}_j \cdot \mathbf{S}_k \right), \\
\label{eq.8.2}
\hat{\bm{\mathcal{I}}}_{ij,k}		&	=	\hat{\dr}_{ij}\frac{24e}{\hbar} \frac{t^3}{U^2} \mathbf{S}_k \cdot \left( \mathbf{S}_i \times \mathbf{S}_j \right),
\end{align}
\end{subequations}
where $\langle ijk \rangle$ denotes an elementary triangle in the lattice, $e$ is the electronic charge, $\hat{\dr}_{ij}$ is the unit vector along the bond $\langle ij \rangle$. The forms of $\delta \hat{\bm{\rho}}_{i,jk}$ and $\hat{\bm{\mathcal{I}}}_{ij,k}$ are uniquely determined by the transformation of these quantities with respect to the following symmetry operations: SU(2) spin rotation, TRS, and inversion operation. 

We now compute the expectation values of the above operators in the spinon ground state. In this regard, we construct the real space spinon Hamiltonian on a finite system of linear size $L = 30$ and obtain the eigenvalues of the corresponding eigenfunctions of the $2L^2 \times 2L^2$ Hamiltonian~\cite{supp}. At first, we rewrite the above operators in Eq.~\eqref{eq.8.1}, and Eq.~\eqref{eq.8.2} in the spinon degrees of freedom using the same mean-field decomposition as in Sec.~\ref{sec:sec_II}. For explicit numerical analysis, we need to fix the mean-field parameters. For subsequent analysis in this section, we work in units of $\tilde{t} = 1$. This leads to a solution of $m_0$ in terms of $\tilde{J}$, and $\tilde{J}_{\chi}$ from Eq.~\eqref{eq.7} as $\tfrac{3m_0}{4} = - \tilde{J}/ \tilde{J}_{\chi} \pm \sqrt{\tilde{J}^2+3 \tilde{J}_{\chi}}/\tilde{J}_{\chi}$. Next, we rewrite Eq.~\eqref{eq.8.1}, and Eq.~\eqref{eq.8.2} in mean-field decomposition as 
\begin{widetext} 
\begin{equation}\label{eq.9}
\delta \hat{\bm{\rho}}_{i,jk} =	\rho_0 \left( e^{i\phi_{ji}}f^{\dagger}_{i} f_{j} + e^{i\phi_{ki}}f^{\dagger}_{i} f_{k} - 2 e^{i\phi_{kj}}f^{\dagger}_{j} f_{k} \right) + {\rm{h.c.}}, \quad
\hat{\bm{\mathcal{I}}}_{ij,k} = \bm{\mathcal{I}}_0 i e^{ i (\phi_{ik} + \phi_{kj}) } f^{\dagger}_{j} f_{i} + {\rm{h.c.}} + \rm{permute} \; \{i,j,k\}, 
\end{equation}
\end{widetext}
where $\rho_0 = e\tfrac{8m_0 t^3}{U^3}$, and $\bm{\mathcal{I}}_0 = \hat{\dr}_{ij} \tfrac{e}{\hbar}\tfrac{9 m_0^2 t^3}{U^2}$ are parameters that depend on the amplitude of the mean-field. Note that we added the contributions of the spin degrees of freedom in Eq.~\eqref{eq.9}~\cite{supp} because of the degenerate spin bands and hence skipped the spin indices. 

To obtain the total charge fluctuation and the loop current for a particular site or a bond, we need to add the contributions of all the shared triangles~\cite{PhysRevB.78.024402,Banerjee2022}. Utilizing the mean-field expressions in Eq.~\eqref{eq.9} for the relevant operators, we calculate their expectation values in the spinon ground state~\cite{supp} for a finite system, as mentioned before. The numerical estimates converge beyond the linear size $L \sim 20 $. In the periodic boundary conditions (PBC), each isolated triangle leads to identical estimates for the charge fluctuation and loop current expectation values. Consequently, there are neither charge redistributions nor circulating loop currents in the cQSL ground state. However, we obtain novel localized charge profiles and loop currents around the system's edges in a finite system \textit{i.e.} with open boundary conditions (OBC). The corresponding results are shown in Fig.~\ref{fig:Fig3}(a). The arrows around the edge signify the magnitude and direction of the localized currents. All values are in units of $2|\bm{\mathcal{I}}_0|$. The magnitude of the loop currents is slightly larger ($\sim 0.7145$) around the corners [$\rm{C}_1$ in Fig.~\ref{fig:Fig3}(a)] which are formed by either an up or down triangle, whereas they are smaller ($\sim 0.6338$) around corners which are composed of both an up and a down triangle [$\rm{C}_2$ in Fig.~\ref{fig:Fig3}(a)]. Note that the loop currents quickly saturate ($\sim 0.6764$) as we move away from the corners along the edges and are consistent with the inversion and $C_6$ rotation symmetries. 

Similarly, a finite charge fluctuation redistributes localized charges around the system's edges, as shown by blue and red circles. In this case, all numbers are shown in the unit of $2\rho_0$. Like the loop currents, the charge profile quickly saturates away from the corners. The maximum charge fluctuations ($+0.0217/-0.0185$) happens around the corner $\rm{C}_1$, whereas the minimum fluctuation ($+0.0201/-0.0163$) occurs around the corners $\rm{C}_2$. The key feature is that the smaller the number of shared triangles for a particular site or a bond, the more the corresponding charge fluctuations or localized currents are, respectively. Most interestingly, the charge separation around edges leads to the formation of a unique dipole moment distribution that can be observed experimentally.

\subsection{Case of a localized spinon \label{sec:sec_III.I}}

The cQSL supports spinons as its low-energy excitation. At the sample edge, there exists a gapless chiral spinon edge mode due to the non-trivial topology of the spinon bands. However, the spinon excitations are gapped inside the bulk. In a clean system with translational invariance in bulk, there are no charge fluctuations or loop currents in bulk [see Fig.~\ref{fig:Fig3}(a)]. Here we focus on an isolated/localized spinon excitation in bulk and discuss its associated electromagnetic responses.

In a clean cQSL, the lower spinon bands with spin up and down are fully occupied. To create a spinon hole, we demand that a specific spin in the spin Hamiltonian does not participate in the fractionalization into spinons. In the mean-field description, this can be achieved by setting the chemical potential for spinons at the pinning site [see Fig.~\ref{fig:Fig3}(b,c)] to be high so that spinons will not occupy the defect site within the low-energy dynamics. This creates a localized spinon hole at the pinning site. Now, we consider a system as before with the defect formed by a large chemical potential at the pinning site as shown in Fig.~\ref{fig:Fig3}(b,c), and impose periodic boundary conditions (PBC). Performing a similar analysis as in Sec.~\ref{sec:sec_III}~\cite{supp}, we notice a redistribution of the charge profile around the localized spinon hole, and a build-up of localized circulating loop current [see Fig.~\ref{fig:Fig3}(b,c)]. As before, all the numbers for charge and current are in units of $2\rho_0$ and $2|\bm{\mathcal{I}}_0|$, respectively. We notice that the circulating loop current around the spinon hole site has the opposite chirality compared to the loop current flowing along the edge [see Fig.~\ref{fig:Fig3}(a)] in the clean system with OBC.
On the other hand, dipole moments formed by the charge redistribution are anti-aligned with the edge dipole moments in the clean system. In the latter case, we only focus on the nearest-neighbor location around the pinning site. Note that the charge profile quickly vanishes away from the pinning center.

\subsection{DMRG calculations \label{sec:sec_III.WZ}}

To validate the above mean-field calculations, we next study the Hamiltonian Eq.~\eqref{eq.2} by using an unbiased DMRG method. Our DMRG calculations focus mainly on the 4-leg cylinder, retaining up to $D = 4000$  U(1) states. We summarize our DMRG results in Fig. \ref{fig:Fig3}(d,e) with $U=10 t$, i.e., deep in the cQSL regime. Here we show the left half of the cylinder for simplicity. We identify that the persistent electric current exists only close to the boundary, manifested by the nontrivial topology and spontaneous TRS breaking of the cQSL phase. The local electrical current quickly reduces from the boundary to the bulk. In the deep bulk, the net current is vanishingly small. 

To create a spinon hole, we can add a local magnetic field $H_{\textrm{loc}} = V_i (n_{i\uparrow}-n_{i\downarrow})$ to the Hamiltonian Eq.~\eqref{eq.2}. (In practice, we add two local magnetic fields and ensure they are separated far away. One pinning point is shown as the green dot in Fig. \ref{fig:Fig3}(e), and the other is in the other half of the cylinder that is not shown here.) The local magnetic field pins the spin locally and forbids it from fractionalizing into delocalized spinons, therefore creating a spinon hole. Around the spinon hole, nonzero electric currents emerge in bulk. Importantly, around the pinned spinon hole, we identify the formation of a loop current (as indicated by the dashed arrow). It is also clear that the electrical charge distribution deviates from the average filling $1$ required for the Mott insulator, as shown in Fig.~\ref{fig:Fig3}(c,e). The general picture of this loop current and charge distribution associated with a spinon hole agrees with the prediction of the mean-field calculations in Sec.~\ref{sec:sec_III.I}. Because of the finite size effect in the narrow direction in DMRG calculations, the current and charge distribution does not respect $C_6$ rotation symmetry along the spinon.

\paragraph*{Numerical estimates:} The DMRG results allow us to estimate the magnitude of the mean-field order parameter $m_0$. Firstly, we provide a rough estimate of $\tilde{J}$, $\tilde{J}_{\chi}$ in Eq.~\eqref{eq.3} based on Refs.~\cite{PhysRevX.10.021042,PhysRevLett.127.087201,supp}. Inserting characteristic values such as $t = 1$ eV, $U = 10$ eV, and $\chi \sim -0.35$~\cite{PhysRevLett.127.087201}, we obtain $\tilde{J} \sim 0.37$ eV and $\tilde{J}_{\chi} \sim 0.15$ eV. Here, $\chi$ is the nonzero chiral order parameter as defined in Fig.~\ref{fig:Fig1}. Since the eigenfunctions of the Hamiltonian in Eq.~\eqref{eq.5} do not depend on the magnitude of $\tilde{t}$, we can compare the loop current magnitudes around the edge of the system obtained by DMRG with our mean-field analysis. Our estimates provide a mean-field amplitude $m_0 \sim 0.1$. Utilizing this in Eq.~\eqref{eq.7}, we obtain an order of magnitude for our phenomenological hopping parameter $\tilde{t} \sim 0.02$ eV. Plugging in the magnitude (obtained by DMRG) of the enclosed loop current around our localized spinons, we estimate an emergent orbital magnetization $\sim 0.01$ $\mu_{\rm{B}}$, where $\mu_{\rm{B}}$ is the Bohr magneton.

\subsection{Quantum field theory description\label{sec:sec_III.QFT}}

The orbital electrical current associated with a spinon can also be understood from the quantum field theory perspective, which sheds further light on the origin of the orbital electrical current. One hallmark of the QSL is the fractionalization of spins and the appearance of an emergent gauge field. Understanding the coupling between the emergent gauge field and the physical electromagnetic fields is crucial for the electromagnetic detection of the QSL. In terms of the parton description, the electron operator can be written as $c_\sigma=b f_\sigma$, where $b$ is a boson operator that carries the electron charge $e$, and $f_\sigma$ is a fermionic spinon operator that carries the spin-$\tfrac{1}{2}$. In cQSL, $f_\sigma$ fermions form Chern bands as was shown in Sec.~\ref{sec:sec_II}. The fractionalization dictates that the charged boson is coupled to both the physical gauge field $\mathbf{A}$ and an emergent gauge field $\mathbf{a}$ as $b\rightarrow b \exp[i (A-a)]$, while the spinon is coupled only to the emergent gauge field as, $f_\sigma \rightarrow f_\sigma \exp(i a)$. The effective low-energy Lagrangian for the $b$ boson has the standard Ginzburg-Landau form (we use the unit $\hbar=e=c=1$)~\cite{Chowdhury2018,PhysRevLett.95.036403,Banerjee2022}
\begin{equation}\label{eqQFT1}
\mathcal{L}_b=\sum_{\mu=x, y}|(i\partial_\mu+a_\mu-A_\mu) b|^2-g |b|^2-\frac{u}{2}|b|^4+\cdots.
\end{equation}
$b$ boson is gapped with $g>0$ in the cQSL which is a Mott insulator. However, there is still a diamagnetic response in $\mathbf{A}-\mathbf{a}$ due to the local current loop in the presence of a magnetic field, similar to Landau diamagnetism in metal, albeit the current loops are strongly localized. Since the $b$ boson is gapped, we can integrate it out to obtain an effective Lagrangian as
\begin{equation}\label{eqQFT2}
\mathcal{L}=\frac{2 C}{4\pi} \epsilon^{\mu\nu\rho} a_\mu\partial_\nu a_\rho-\frac{\chi_b}{2}[\nabla\times(\mathbf{a}-\mathbf{A})]^2-\frac{\chi_B}{2}(\nabla\times \mathbf{A})^2,
\end{equation}
where the first term on the right-hand side is the Chern-Simon term obtained by integrating out  $f_\sigma$ that fills topological Chern bands with a Chern number $C$ ($C=1$ in our model). Here $\chi_b$ accounts for the diamagnetic susceptibility due to the gapped boson $b$, $\chi_B$ is the susceptibility of the background~\cite{PhysRevB.97.045152}. It is clear from the Chern-Simon term that a spinon carries $\pi/C$ flux of $\mathbf{a}$~\cite{Lin_2022}. The physical magnetic field associated with the emergent magnetic field, which can be seen from Eq.~\eqref{eqQFT2} by minimizing $\mathcal{L}$ with respect to $\mathbf{B}\equiv \nabla\times \mathbf{A}$, is: $\mathbf{B}=\chi_b/(\chi_b+\chi_B)\nabla\times\mathbf{a}$. Hence a spinon excitation induces an orbital electrical current with a total flux of $\mathbf{B}$ equal to $\chi_b\pi/(\chi_b+\chi_B)C$.

\begin{figure*}[t]
\centering
\includegraphics[width=1\linewidth]{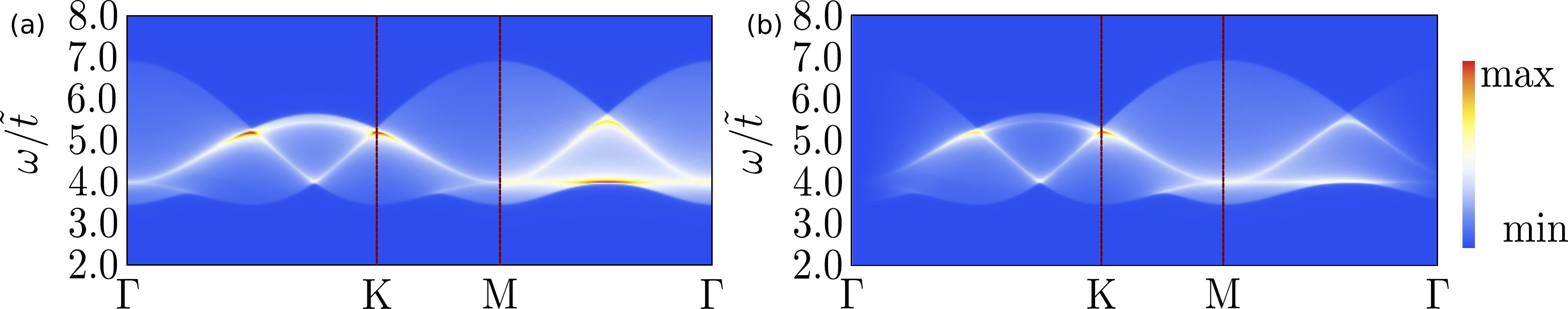} 
\caption{The normalized scattering density of states $\sf{g}(\vq, \omega)$ (a) (see definition in the main text) and the dynamic spin structure factor $S(\vq,\omega)$ (b) along the high symmetry points of the triangular lattice Brillouin zone (BZ). The analysis is performed neglecting any spinon interactions which are present when the dynamics of the emergent gauge field are considered explicitly.}\label{fig:Fig4}
\end{figure*} 

\subsection{Dynamic spin-structure factor \label{sec:sec_III.II}}

In the previous sections, we established that spinon excitations in the cQSL phase carry orbital electrical loop currents and charges. Now, we proceed to investigate the electromagnetic response of a cQSL in terms of optical conductivity and Faraday rotation. Before considering the optical conductivity, which involves higher-order spinon correlation functions, we consider the standard dynamic spin-structure factor (DSSF) in the framework of spinon description. DSSF is an essential physical quantity that is routinely used as an experimental tool to probe the nature of the magnetic ground state and is defined as
\begin{equation}\label{eq.10}
S(\vq,\omega) 
=
\sum_{i,j} \frac{e^{i\vq \cdot(\dr_i - \dr_j)}}{N_{s}} \int_{-\infty}^{\infty} dt e^{i\omega t } \braket{\mathbf{S}_{i}(t) \cdot \mathbf{S}_j(0)},
\end{equation}
where $N_{s}$ denotes the number of sites, and $\vq, \omega$ denotes the probe momentum and frequency, respectively. With the two-sublattice structure as illustrated in Fig.~\ref{fig:Fig2}(a), we first rewrite Eq.~\eqref{eq.10} in terms of spinon operators. The above expression simplifies upon utilizing the spectral representation with the weighted summation over the sub-lattice resolved spin-structure factors. The latter is written as~\cite{supp}
\begin{widetext}
\begin{equation}\label{eq.11}
S^{\eta \zeta}_{}(\vq,\omega)
=
\frac{3}{2} \sum_{n,\vk}
\braket{0|f^{\dag}_{\eta,\vk} f_{\eta,\vk+\vq}|n} \braket{n|f^{\dag}_{\zeta,\vk+\vq} f_{\zeta,\vk}|0} \delta(\omega - E_n + E_0), \quad
\{ \eta,\zeta \} \in  {\rm{A}, \rm{B}}
\end{equation}
\end{widetext}
where $\{ \rm{A}, \rm{B} \}$ denotes the two sublattice degrees of freedom, and $E_n$ denotes the eigen energy of the $n$ -th excited state. Note that we added the contributions from the degenerate spin up and down bands, and consequently skipped the indices as before. Rewriting in the diagonal basis and summing the sublattice degrees of freedom, we obtain the DSSF in our phenomenological cQSL. In Fig.~\ref{fig:Fig4}(b), we show the DSSF profile. Note that we’ve adopted a normalization where the absolute maximum is set to unity for convenience. The excited state $\ket{n}$ contains one pair of spinon hole and spinon excitation, or spinon exciton, as evident from Eq.~\eqref{eq.11}. 

We notice that apart from a relatively strong peak centered in a narrow region around the edge of the BZ at the $K$ point, there are almost no sharp features within the BZ. The broad continuum in the BZ reflects the absence of any long-range magnetic order, i.e., there are no well-defined magnon excitations at a given momentum $\vq$ with energy $\omega$. The relatively broad/diffused bands (illustrated by the white halos)  correspond to a two-spinon continuum. At $\vq =0$, the DSSF corresponds to the vertical spinon exciton, as is evident from Eq.~\eqref{eq.11}. In this case, the wave function overlap between the wave function of the spinon hole in the occupied band and the spinon in the unoccupied band is zero at the same momentum and subsequently leads to a vanishing weight distribution around $\Gamma$ point as seen in Fig.~\ref{fig:Fig4}(b). To illustrate this, we also plot the scattering density of states $\sf{g}(\omega,\vq) = \sum_\vk \delta(\omega - \varepsilon_{\vk+\vq} - \vq)$ in Fig.~\ref{fig:Fig4}(a), where there is a finite spectral weight around the $\Gamma$ point. The absence of spectral weight around the $\Gamma$ point is common to the cQSL phase in other lattices, \textit{viz}. kagome~\cite{Punk2014,PhysRevB.94.104413}. In reality, the fluctuations of the emergent gauge field around the mean-field saddle point mediate the attraction between the spinon hole and the spinon, which has been neglected in the present discussion. However, even in this case, the spectral weight around $\Gamma$ point will vanish due to the zero overlap of the eigenfunctions~\cite{Punk2014}.

\subsection{Optical conductivity and Faraday rotation \label{sec:sec_III.III}}

Finally, we focus on the main result of our work by showing that optical responses below the Mott gap can be used to probe the emergent cQSL state in the TLHM~\cite{PhysRevB.87.245106,PhysRevB.90.121105,Hwang_2014}. The longitudinal and the transverse optical conductivity in this regime become nonvanishing because of the finite electronic polarization. Following the work by Bulaevskii et al. ~\cite{PhysRevB.78.024402}, we obtain the corresponding expression for a three-site problem as 
\begin{subequations}
\begin{align}
\label{eq.12.1}
P_x 	&	=	4\sqrt{3} ea \frac{t^3}{U^3} \left( \mathbf{S}_i \cdot \mathbf{S}_j + \mathbf{S}_i \cdot \mathbf{S}_k - 2 \mathbf{S}_j \cdot \mathbf{S}_k \right), \\
\label{eq.12.2}
P_y 	&	=	12 ea \frac{t^3}{U^3} \left( \mathbf{S}_i \cdot \mathbf{S}_j - \mathbf{S}_i \cdot \mathbf{S}_k \right),
\end{align}
\end{subequations}
where $a$ is the lattice constant, and $t,U$ are the parameters defined as before in Eq.~\eqref{eq.2}. The above two expressions are particularly relevant as we deal with a triangle lattice. However, note that within a lattice framework, we need to add the contributions of all the triangles surrounding a particular site $i$ to obtain the total polarization $\mathbf{P}$. The latter naturally couples to an external electric field as $-\mathbf{P} \cdot \mathbf{E}(t)$. Consequently, the associated optical conductivity within the linear response theory reads~\cite{Mahan2011,Marder2010,Hwang_2014} 
\begin{equation}\label{eq.13}
\sigma_{ab} (\omega)
=
\frac{i\omega}{V \hbar} 
\sum_{n \neq 0} \frac{\braket{\psi_0 | P_a | \psi_n} \braket{\psi_n | P_b | \psi_0}}{\omega - \omega_n + i \epsilon} 
+
\substack{a \leftrightarrow b \\ \omega_n \rightarrow - \omega_n} ,  
\end{equation}
where $\ket{\psi_0}$, and $\ket{\psi_n}$ are the ground and excited states, respectively, $\mathcal{H} \ket{\psi_n} = E_n \ket{\psi_n} \forall n \in \{0,1,2, \ldots \}$, $V$ is the volume, and $\omega_n = E_n -E_0$, where $E_0$ is the energy of the ground state. Note that the above expression is valid in the frequency regime much less than the energy scale ($U$) associated with the charge gap in the Hubbard model, i.e., $\hbar \omega \ll U$. Additionally, broken TRS in the chiral phase immediately implies non-vanishing off-diagonal components ($a \neq b$). This leads to a finite MOFE signal proportional to the real part of the transverse optical conductivity defined in Eq.~\eqref{eq.1}. 

\begin{figure}[b]
\centering
\includegraphics[width=1.0\linewidth]{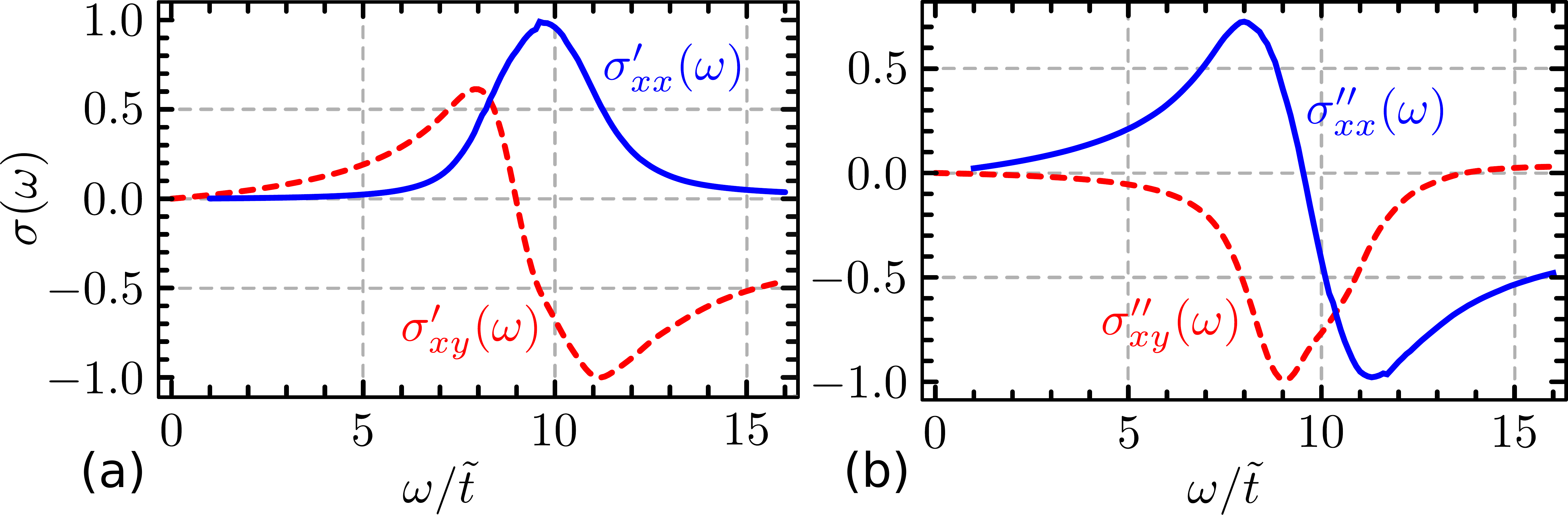} 
\caption{The real (a) and the imaginary (b) part of the normalized transverse (\textit{dashed line}) and longitudinal (\textit{solid line}) optical conductivity as a function of the frequency of the incident light.}\label{fig:Fig5}
\end{figure} 

We proceed as before in Sec.~\ref{sec:sec_III.II} by rewriting the polarization operator in terms of spinon degrees of freedom. Readers are referred to Ref.~\cite{supp} for the details of the calculations. However, in stark contrast to the DSSF analysis, here we need to consider the correlation functions involving eight spinon operators as is evident from Eq.~\eqref{eq.13}~\cite{supp}. We perform numerical integration in Mathematica with a quasi-Monte Carlo routine and obtain the transverse and longitudinal optical conductivity as a function of the frequency as shown in Fig.~\ref{fig:Fig5}. Both the real ($\sigma'$) and imaginary ($\sigma''$) part of the quantities are shown in panel (a) and panel (b), respectively. Similar to Sec.~\ref{sec:sec_III.II}, we adopted a normalization in which the absolute maximum of the quantities is set to unity.

We notice that $\sigma'_{xy}(\omega)$ changes sign at a frequency $\omega_0 \sim 9\tilde{t}$ that is almost twice the spinon gap around the BZ edge at the $M$ point. Around the same frequency $|\sigma''_{xy}(\omega)|$ attains its largest magnitude. $\sigma'_{xx}(\omega)$, and $\sigma''_{xx}(\omega)$ also show similar characteristics at frequencies close to twice the spinon gap at the $M$ point. Plugging in characteristic numbers as $t = 1$ eV, $U = 10$ eV, $m_0 \sim 0.1$, $a \sim 10$ \AA, and $\tilde{t} \sim 0.02$ eV, we obtain $\sigma'_{xy} \sim 3 \times 10^{-6} \tfrac{e^2}{\hbar}$ for $\omega \sim 20$ THz. This leads to an estimated Faraday rotation angle of around $0.2$ mRad/$\mu\rm{m}$ per thickness of the sample. The magnitude is within the allowed resolution of current experiments~\cite{PhysRevB.86.235133}.

\section{Discussion and Concluding Remarks \label{sec:sec_IV}}

This paper provides extensive mean-field analysis for the electromagnetic response of a cQSL phase. We started from a phenomenological cQSL Hamiltonian as in Eq.~\eqref{eq.3} and analyzed the spectrum of fractionalized excitations in terms of spinon mead field theory. Despite being deep inside the Mott insulator regime, where the charge degrees of freedom are gapped, we obtain a nonvanishing electrical loop current distribution and charge fluctuations associated with a localized spinon excitation. Additionally, we performed unbiased DMRG calculations in the triangular lattice Hubbard model at the intermediate coupling regime, where the cQSL is stabilized. The DMRG results confirm the physical picture of the parton mean-field results, where both approaches provide similar structures of the loop currents and charge redistributions in the cQSL phase, as illustrated in Fig.~\ref{fig:Fig3}. The DMRG calculations further allow us to estimate the magnitude of the electrical charge and orbital current associated with a spinon excitation. Assuming a typical value of $t = 1$ eV and $U = 10$ eV, we estimate the electrical current and charge around the localized spinons to be around 17 $\mu$A, and $\pm 0.1\%$ of $e$, respectively. In addition, we performed quantum field theory analysis to unravel the connection between the spinon excitation and emergent and physical gauge fields, which clearly shows that a flux of the physical magnetic field dresses a localized spinon.

The electromagnetic characteristics of spinon excitations immediately imply a nonvanishing optical response in the cQSL. We compute the optical response functions by focusing on the optical conductivity. The nonvanishing transverse optical conductivity $\sigma'_{xy}(\omega)$ below the Mott gap can be considered a smoking gun signature of the underlying chiral nature of the QSL. Since a finite $\sigma'_{xy}(\omega)$ signifies a non-zero Faraday rotation angle $\Theta_{\rm{F}}$, our predictions can be directly tested by suitable optical techniques such as MOFE or Kerr effect. Since $\sigma'_{xy}(\omega)$ changes sign as the frequency increases, an experimental signature of cQSL would be to see if, as a function of incoming photon frequency, the Faraday rotation angle changes sign or not. For completeness and as an intermediate step, we also analyze the dynamic spin-structure factor of the cQSL as illustrated in Fig.~\ref{fig:Fig4}(b). The absence of sharp features signifies no well-defined magnon excitations in the QSL. 

In cQSL, each unit triangle carries an orbital current. However, this orbital current cancels in the bond shared by two neighboring triangles for a translationally invariant system. This cancelation is not perfect in the presence of impurities or near edges, leaving finite orbital magnetization localized around impurities. Therefore, the orbital magnetization localized around impurities already serves as a signature of time-reversal symmetry breaking in QSL. This defect-induced orbital magnetization can be distinguished from spinons, which are dynamical excitations (despite being gapped) of cQSL. Depending on the protocol to tune the system into the cQSL, spinons can be created at different system locations, and the protocol can control their density. In contrast, the orbital magnetization localized around impurities does not depend on the protocol. 

Compared to our previous theoretical work on Kitaev materials~\cite{Banerjee2022}, here, TRS is spontaneously broken due to considerable charge fluctuations in a Hubbard model at intermediate coupling strength. As noted in our quantitative estimates for the loop current or associated charge polarization, the latter translates into a larger electromagnetic response. Note that the associated gauge structure for the cQSL in the TLHM is U(1), whereas the Kitaev spin liquid has a $Z_2$ gauge structure.

In summary, we show that spinon excitations in cQSL carry an electrical charge and orbital current, despite the system being a Mott insulator. Such an electromagnetic response can be detected experimentally using the MOFE or Kerr effect. Therefore, our work provides a clear electromagnetic signature of the cQSL, which helps determine the nature of nonmagnetic states observed in certain materials realizing the triangular lattice Mott insulator. 
 
\section{Acknowledgement \label{sec:sec_V}}

The authors thank Vivien Zapf and S. S. Gong for helpful discussions. This work was carried out under the auspices of the US DOE NNSA under Contract No. 89233218CNA000001 through the LDRD Program, and was performed, in part, at the Center for Integrated Nanotechnologies, an Office of Science User Facility operated for the U.S. DOE Office of Science, under user proposals \#2018BU0010 and \#2018BU0083. The computational part of the program was supported by "Pioneer" and "Leading Goose" R\&D Program of Zhejiang (2022SDXHDX0005), the Key R\&D Program of Zhejiang Province (2021C01002), National Key R\&D Program (2022YFA1402200). We thank Westlake University HPC Center for computation support.

\bibliographystyle{apsrev4-1}
\bibliography{Ms_csl}


\widetext
\clearpage
\begin{center}
\large{\textbf{Supplementary material:\--- \\ Electromagnetic signatures of chiral quantum spin liquid}} \\
\end{center}

\begin{center}
\text{Saikat Banerjee$^1$, Wei Zhu$^{2,3}$, and Shi-Zeng Lin$^{4,5}$}
\end{center}
\begin{center}
\textit{
$^{1}$Theoretical Division, T-4, Los Alamos National Laboratory, Los Alamos, New Mexico 87545, USA \\
$^2$School of Science, Westlake University, No. 600 Dunyu Road, Hangzhou 310030, China \\
$^3$Key Laboratory for Quantum Materials of Zhejiang Province, Westlake University, Hangzhou 310024, China \\
$^4$Theoretical Division, T-4 and CNLS, Los Alamos National Laboratory, Los Alamos, New Mexico 87545, USA \\
$^5$Center for Integrated Nanotechnologies (CINT), \\ Los Alamos National Laboratory, Los Alamos, New Mexico 87545, USA}
\end{center}

\setcounter{equation}{0}
\setcounter{section}{0}
\setcounter{figure}{0}
\setcounter{table}{0}
\setcounter{page}{1}

\makeatletter

\renewcommand{\theequation}{S\arabic{equation}}
\renewcommand{\thefigure}{S\arabic{figure}}
\renewcommand{\bibnumfmt}[1]{[S#1]}
\renewcommand{\citenumfont}[1]{S#1}


\onecolumngrid
\clearpage

\section{Hubbard model to chiral spin model \label{sec:1}}

In this section, we outline the steps for phenomenologically obtaining the chiral spin model starting from a Hubbard model on a triangular lattice at half-filling. The corresponding Hamiltonian is written as
\begin{equation}\label{Eq.1}
\mathcal{H} = -t \sum_{\langle ij \rangle,\sigma} c^{\dagger}_{i\sigma} c_{j\sigma} + U\sum_{i} n_{i\uparrow}n_{i\downarrow},
\end{equation} 
where $\langle ij \rangle$ corresponds to the nearest-neighbor tight-binding model on a triangular lattice, $\sigma$ corresponds to the spin degrees of freedom, and $U$ is the strength of the local Hubbard repulsion. In this case, the low-energy effective spin Hamiltonian in the strong-coupling limit ($U \gg t$) can be obtained through Schrieffer-Wolff transformation (SWT)~\cite{Paper1,Paper2,Paper3} as
\begin{subequations}
\begin{align}
\label{Eq.2.1}
\mathcal{H}_{\rm{eff}} & = \mathcal{H}^{(2)}_{\rm{eff}} + \mathcal{H}^{(4)}_{\rm{eff}}, \quad \mathcal{H}^{(2)}_{\rm{eff}} 
= 
J^{(2)} \sum_{\langle ij \rangle} \mathbf{S}_i \cdot \mathbf{S}_j, \\
\label{Eq.2.2}
\mathcal{H}^{(4)}_{\rm{eff}} &
= 
J^{(4)}_1 \sum_{\langle ij \rangle} \mathbf{S}_i \cdot \mathbf{S}_j + 
J^{(4)}_2 \sum_{\llangle ij \rrangle} \mathbf{S}_i \cdot \mathbf{S}_j +
J^{(4)}_3 \sum_{\langle\llangle ij \rrangle \rangle} \mathbf{S}_i \cdot \mathbf{S}_j +
J^{(4)}_{\mathrm{R}} \sum_{\langle i,j,k,l \rangle} \mathcal{R}_{ijkl}, \\
\label{Eq.2.3}
\mathcal{R}_{ijkl} &
= 
(\mathbf{S}_i \cdot \mathbf{S}_j) (\mathbf{S}_k \cdot \mathbf{S}_l) + (\mathbf{S}_i \cdot \mathbf{S}_l) (\mathbf{S}_j \cdot \mathbf{S}_k) - (\mathbf{S}_i \cdot \mathbf{S}_k) (\mathbf{S}_j \cdot \mathbf{S}_l),
\end{align}
\end{subequations}
where the exchange couplings are given by~\cite{Paper3}
\begin{equation}\label{Eq.3}
J^{(2)} 	= 	\frac{4t^2}{U}, 				\quad
J^{(4)}_1 	= 	-\frac{24t^4}{U^3},				\quad 
J^{(4)}_2   =	J^{(4)}_3 = \frac{4t^4}{U^3},	\quad 
J^{(4)}_{\mathrm{R}} 	  = \frac{80t^4}{U^3}.
\end{equation}
In a previous theoretical work~\cite{Paper4}, it was shown that the ring exchange term leads to an induced chirality in the low-energy spin dynamics of the TLHM. It is worth mentioning that such a flux phase was previously pointed out in triangular lattice material $\kappa$-(ET)$_2$Cu$_2$(CN)$_3$ by Motrunich within the mean-field description in Ref.~\cite{Paper5}. Motivated by these studies, we consider the following chiral spin liquid model as
\begin{equation}\label{Eq.4}
\mathcal{H}_{\rm{pheno}} 
= 
\tilde{J} \sum_{\langle ij \rangle} \mathbf{S}_i \cdot \mathbf{S}_j 
+ 
\tilde{J}_{\chi} \sum_{\substack{\llangle ijk \rrangle \\ \bigtriangleup,\bigtriangledown}} \mathbf{S}_i \cdot (\mathbf{S}_j \times \mathbf{S}_k),
\end{equation}
where $\tilde{J}$, and $\tilde{J}_{\chi}$ are provided in Ref.~\cite{Paper4} as
\begin{equation}\label{Eq.5}
\tilde{J} 			= J^{(2)} - \frac{107}{88} J_{\rm{R}}^{(4)}, \quad
\tilde{J}_{\chi} 	= 3 J_{\rm{R}}^{(4)} \left( - \frac{39}{11} \chi - \frac{1344}{11} \chi^3 + \frac{12288}{11} \chi^5 \right),
\end{equation}
where $\chi = \braket{\mathbf{S}_i \cdot (\mathbf{S}_j \times \mathbf{S}_k}$ is the non-vanishing chiral order parameter in the emergent chiral QSL state. Self-consistent density-matrix renormalization analysis in Ref.~\cite{Paper4,Paper6} shows that it is non-vanishing in a wide region between $U_{\rm{c}_1} \sim 9 U$ and $U_{\rm{c}_2} \sim 10.75 U$. Therefore, the parameters of the phenomenological Hamiltonian in the main text are directly related to the parameters of the original Hubbard model.

\section{Phenomenological description \label{sec:sec.2}}

We assume that the underlying fractionalized excitations are spinons, and correspondingly rewrite the spin degrees of freedom as~\cite{Paper7} $\mathbf{S}_i = \tfrac{1}{2} f^{\dagger}_{i\alpha} \bm{\sigma}_{\alpha \beta} f_{i\beta}$. Here, $f^{\dag}_{i\alpha}$ creates a spinon at site $i$ with spin $\alpha$, and $\bm{\sigma}$ denotes the vector of the Pauli matrices. We utilize the product relation for the Pauli matrices as
\begin{equation}\label{Eq.6}
\bm{\sigma}_{\alpha \beta} \cdot \bm{\sigma}_{\gamma \delta} = 2 \delta_{\alpha \delta} \delta_{\beta \gamma} - \delta_{\alpha \beta} \delta_{\gamma \delta}.
\end{equation}
Plugging this back into Eq.~(3) of the main text, we first obtain the Heisenberg part as 
\begin{equation}\label{Eq.7}
\mathcal{H}_{\mathrm{heisen}} = -\frac{\tilde{J}}{2} \sum_{\langle ij \rangle} f^{\dagger}_{i\alpha} f_{j\alpha} f^{\dagger}_{j\beta} f_{i \beta} - \frac{\tilde{J}}{4} \sum_{\langle ij \rangle} n_{i} n_{j},
\end{equation}
where $n_i = f^{\dagger}_{i\alpha} f_{i\alpha}$ and we assume the summation over the repeated indices unless explicitly mentioned. We note that the spinon description manifestly enlarges the physical Hilbert space. To remain in the physical Hilbert space, we utilize the half-filling constraint per site as $\sum_{\alpha} f^{\dagger}_{i\alpha} f_{i\alpha} = 1$. However, in this work, we assume this constraint to be loosely applicable in an average way in the spirit of mean field theory. Next, we perform only particle-hole mean-field decomposition to rewrite the Hamiltonian in Eq.~\eqref{Eq.7} as 
\begin{equation}\label{Eq.8}
\mathcal{H}_{\mathrm{heisen}} = -\frac{\tilde{J}}{2} \sum_{\langle ij \rangle} \left( m_{ji} f^{\dagger}_{i\alpha} f_{j\alpha} + \mathrm{h.c.} \right) + \frac{\tilde{J}}{2} \sum_{\langle ij \rangle} |m_{ij}|^2,
\end{equation}
where we have ignored the last term in Eq.~\eqref{Eq.7}, and introduced a mean-field ansatz as $m_{ij} = \langle f^{\dagger}_{i\alpha} f_{j \alpha} \rangle$. Similarly, we rewrite the chiral term as~\cite{Paper8}
\begin{align}
\nonumber
\mathcal{H}_{\mathrm{chiral}} 
& 
= \tilde{J}_{\chi} \sum_{\llangle ijk \rrangle} \mathbf{S}_i \cdot \left( \mathbf{S}_j \times \mathbf{S}_k \right) \\
\nonumber
& 
= \frac{J_{\chi}}{8} \sum_{\llangle ijk \rrangle} 
\left( f^{\dagger}_{i \alpha} \bm{\sigma}_{\alpha \beta} f_{i \beta} \right) \cdot 
\left( f^{\dagger}_{j \alpha^{'}} \bm{\sigma}_{\alpha^{'} \beta^{'}} f_{j \beta^{'}} \right) \times
\left( f^{\dagger}_{k \alpha^{''}} \bm{\sigma}_{\alpha^{''} \beta^{''}} f_{k \beta^{''}} \right) \\
\nonumber
& = \frac{J_{\chi}}{8} \sum_{\llangle ijk \rrangle} 
\epsilon_{abc} (\sigma^a)_{\alpha \beta} (\sigma^b)_{\alpha^{'} \beta^{'}} (\sigma^c)_{\alpha^{''} \beta^{''}} f^{\dagger}_{i \alpha} f_{i \beta} f^{\dagger}_{j \alpha^{'}} f_{j \beta^{'}} f^{\dagger}_{k \alpha^{''}} f_{k \beta^{''}} \\
\label{Eq.9}
& \approx
\frac{3i \tilde{J}_{\chi}}{16} \sum_{\llangle ijk \rrangle} 
\left( 
-m_{ik} m_{kj} m_{ji} 
+ m_{kj} m_{ji} f^{\dagger}_{i\alpha} f_{k \alpha}
+ m_{ik} m_{kj} f^{\dagger}_{j\alpha} f_{i \alpha}
+ m_{ji} m_{ik} f^{\dagger}_{k\alpha} f_{j \alpha}
- \mathrm{h.c.} \right),
\end{align}
The total Hamiltonian is then $\mathcal{H}_{\rm{pheno}} = \mathcal{H}_{\mathrm{heisen}} + \mathcal{H}_{\mathrm{chiral}}$. Without going into a self-consistent mean-field analysis, we assume a particular form of the mean fields and benchmark our analysis with our unbiased DMRG calculations. Assuming translational invariance, we choose $m_{ij} = m_0 e^{i\phi_{ij}}$, where $m_0$ is the amplitude of the order parameter and $\phi_{ij}$ are the bond-dependent phases. Now ignoring the amplitude and phase fluctuations, we can write the Hamiltonian as 
\begin{equation}\label{Eq.10}
\mathcal{H}_{\mathrm{pheno}} 
=
-\frac{\tilde{J} m_0}{2} \sum_{\langle ij \rangle} e^{i\phi_{ji}} f^{\dagger}_{i\alpha} f_{j \alpha} +
\frac{3i \tilde{J}_{\chi} m_0^2}{16} \sum_{\langle ij \rangle}  e^{ i (\phi_{ik} + \phi_{kj}) } f^{\dagger}_{j\alpha} f_{i \alpha} + 
\{ i\leftrightarrow j \leftrightarrow k\} + 
3\tilde{J} \sum_{ i} m^2_0 + 
\frac{9\tilde{J}_{\chi}}{8} \sum_{i} m_0^3 \cos \theta + \mathrm{h.c}
\end{equation}
Combining the hopping phases $\phi_{ij}$'s, the above Hamiltonian can be written in a compact form as $\mathcal{H}_{\mathrm{pheno}} = - \tilde{t} \sum_{\langle ij \rangle} e^{i\psi_{ij}} f^{\dagger}_{i \alpha} f_{j \alpha} + \mathrm{h.c.}$, where $\tilde{t}$ is the spinon hopping amplitude and is related to our phenomenological parameters as
\begin{subequations}
\begin{align}
\label{Eq.11.1}
\tilde{t} \cos \psi_{ij} & = \frac{\tilde{J} m_0}{2} \cos \phi_{ji} + \frac{3 \tilde{J}_{\chi} m_0^2}{16} \sin \left( \phi_{ik} + \phi_{kj} \right), \\
\label{Eq.11.2} 
\tilde{t} \sin \psi_{ij} & = \frac{\tilde{J} m_0}{2} \sin \phi_{ji} + \frac{3 \tilde{J}_{\chi} m_0^2}{16} \cos \left( \phi_{ik} + \phi_{kj} \right),
\end{align}	
\end{subequations}
where $\psi_{ij} + \psi_{jk} + \psi_{ki} = \Phi_0$ with $\Phi_0$ being total flux enclosed within a single triangular plaquette. At this point, all $\phi_{ij}/\psi_{ij}$ remains undetermined. Now we utilize Lieb's theorem~\cite{Paper9} to determine the phases. According to the theorem, a fermion hopping on a bipartite lattice realizes the ground state with $\pi$-flux square plaquettes. Consequently, we consider a doubled unit cell such that one up and one down triangle jointly form the rhombus-like bipartite unit cell as shown in Fig.~1(a), and impose a $\pi$-flux in the doubled unit cell. This still leaves us with various choices for the bond-dependent phases $\psi_{ij}$, $\psi_{jk}$, and $\psi_{ki}$ forming the triangular plaquette $\braket{ijk}$. A generic choice of the bond-dependent phases is shown in Fig.~1(a), where $\theta$ is some arbitrary angle specifying whether both the up and down triangles have the same flux $\pi/2$ or some staggered flux configurations as $\pi/2 \pm \theta$ (both adding to $\pi$ in the rhombus-shaped unit cell). 

\subsection{Topological spinon bands \label{sec:sec.3}}

\begin{figure}[t]
\centering
\includegraphics[width=0.7\linewidth]{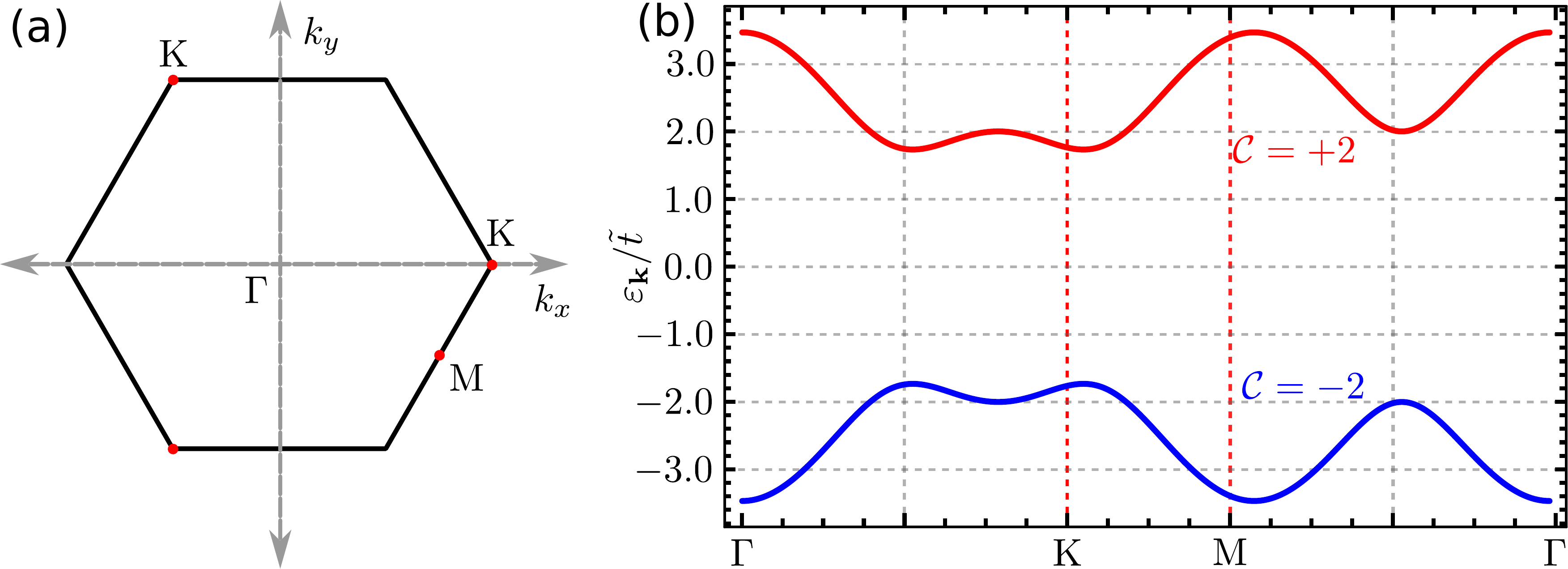} 
\caption{(a) The Brillouin zone of a triangular lattice with the high-symmetry points marked by the filled circles. (b) The spectrum of gapped spinons within our mean-field approximation with the staggered phase $\theta = 0$. The topological Chern numbers are labeled on the two bands. Each spinon band is doubly degenerate in the spin degrees of freedom.}\label{fig:SFig1}
\end{figure}

We now move on to compute the spinon bands within the phenomenological flux phases in the triangle lattice. First of all, the primitive and the reciprocal lattice vectors of the original triangular lattice are given by
\begin{equation}\label{Eq.12}
\mathbf{a}_1 = a \left( 1,\; 0 \right), \; \; \mathbf{a}_2 = \frac{a}{2} \left(1,\; \sqrt{3} \right), 
\qquad
\mathbf{b}_1 = \frac{2\pi}{\sqrt{3}a} \left( \sqrt3{},\; -1 \right), \; \; \mathbf{b}_2  = \frac{4\pi}{\sqrt{3}a} \left(0,\; 1 \right),
\end{equation}
where $a$ is the lattice constant. The corresponding nearest-neighbor vectors as given by
\begin{equation}\label{Eq.13}
\bm{\delta}_1 = a \left(1, \; 0 \right), 
\quad
\bm{\delta}_2 = \frac{a}{2} \left(1, \; \sqrt{3} \right), 
\quad
\bm{\delta}_3 = \frac{a}{2} \left(1, \; - \sqrt{3} \right).
\end{equation}
We also show the corresponding Brillouin zone (BZ) in Fig.~\ref{fig:SFig1}(a) with the high-symmetry points as 
\begin{equation}\label{Eq.13.1}
\Gamma = (0, \; 0), 
\quad 
\mathrm{K} = \frac{\pi}{a} \left( \frac{4}{3}, \; 0 \right),
\quad 
\mathrm{M} = \frac{\pi}{a} \left( 1, \; \frac{1}{\sqrt{3}} \right).
\end{equation}
Since the previous flux configuration doubles the unit cell as $\mathbf{a}_1 \rightarrow 2 \mathbf{a}_1$, the tight-binding Hamiltonian in the sub-lattice basis (see Fig.~1(a) in the main text) is written as (note that we consider the uniform flux configuration with $\theta = 0$)
\begin{equation}\label{Eq.14}
\mathcal{H}_{\rm{eff}} 
=   
-\tilde{t} \sum_{i} 
\left(
f^{\dagger}_{i\mathrm{A}} f_{i + \bm{\delta}_1 \mathrm{B}} + 
f^{\dagger}_{i\mathrm{A}} f_{i + \bm{\delta}_2\mathrm{A}} +
i f^{\dagger}_{i\mathrm{A}} f_{i + \bm{\delta}_3\mathrm{B}} +
f^{\dagger}_{i\mathrm{B}} f_{i + \bm{\delta}_1\mathrm{A}} -
f^{\dagger}_{i\mathrm{B}} f_{i + \bm{\delta}_2\mathrm{B}} -
i f^{\dagger}_{i\mathrm{B}} f_{i + \bm{\delta}_3\mathrm{A}} \right) + \mathrm{h.c.},
\end{equation}
Translating into the momentum space, we obtain
\begin{equation}\label{Eq.15}
\mathcal{H}_{\rm{eff}} 
=
- 2\tilde{t} \sum_{\vk}
\begin{pmatrix}
f^{\dagger}_{\vk \mathrm{A}} & f^{\dagger}_{\vk \mathrm{B}}
\end{pmatrix}
\begin{pmatrix}
\cos \vk \cdot \bm{\delta}_2 									&		\cos \vk \cdot \bm{\delta}_1 + i\cos \vk \cdot \bm{\delta}_3 \\
\cos \vk \cdot \bm{\delta}_1 - i\cos \vk \cdot \bm{\delta}_3 	& 		-\cos \vk \cdot \bm{\delta}_2 
\end{pmatrix}
\begin{pmatrix}
f_{\vk \mathrm{A}} \\
f_{\vk \mathrm{B}}
\end{pmatrix}.
\end{equation}
The above Hamiltonian can be written in a compact form with Pauli matrices as $\mathcal{H}_{\rm{eff}} = - 2t \sum_{\vk} \mathbf{d}_{\vk} \cdot \bm{\sigma}$ where
\begin{equation}\label{Eq.17}
d_{1\vk} = \cos \vk \cdot \bm{\delta}_2, 
\quad
d_{2\vk} = \cos \vk \cdot \bm{\delta}_1, 
\quad
d_{3\vk} = \cos \vk \cdot \bm{\delta}_3.
\end{equation}
The gapped spinon spectrum is obtained by diagonalizing the Hamiltonian in Eq.~\eqref{Eq.15}. The dispersion is given by $\varepsilon_{\vk} = \pm 2 \tilde{t} \sqrt{d_{1\vk}^2 + d_{2\vk}^2 + d_{3\vk}^2}$ (see Fig.~1(b) in the main text). The spectrum remains gapped for any other choice of $\theta$, except at $\theta = \tfrac{\pi}{2}$ when the gap closes as depicted in Fig.~1(c) in the main text. We computed the Chern number in the gapped phase using link variable method~\cite{Paper10}, and find the Chern numbers for the bands to be $\pm 2$ (upon adding the spin degenerate bands) [see Fig.~\ref{fig:SFig1}(b)]. 

\subsection{Orbital electrical current and charge fluctuation in the gapped phase \label{sec:sec.4}}

In this section, we first provide the steps leading to an emergent non-vanishing loop electrical current distribution in the CSL phase. The current operator in the single-band Hubbard model reads as~\cite{Paper11}
\begin{equation}\label{Eq.18}
\hat{\bm{\mathcal{I}}}_{ij,k} = \hat{\dr}_{ij}\frac{24e}{\hbar} \frac{t^3}{U^2} \mathbf{S}_k \cdot \left( \mathbf{S}_i \times \mathbf{S}_j \right),
\end{equation}
where $\hat{\dr}_{ij}$ is the unit vector connecting two sites $i,j$, and the localized current flows within a triangular loop. Rewriting in terms of the spinons, we obtain 
\begin{align}
\nonumber
\hat{\bm{\mathcal{I}}}_{ij,k} =
&
\hat{\dr}_{ij} \frac{e}{\hbar} \frac{9 i t^3}{2U^2} \left(
m_{kj} m_{ji} f^{\dagger}_{i\alpha} f_{k \alpha}
+ m_{ik} m_{kj} f^{\dagger}_{j\alpha} f_{i \alpha}
+ m_{ji} m_{ik} f^{\dagger}_{k\alpha} f_{j \alpha}
- \mathrm{h.c.}
\right) \\
\label{Eq.19}
&
=
\hat{\dr}_{ij} \frac{9 i m_0^2 t^3}{U^2} 
e^{ i (\phi_{ik} + \phi_{kj}) } f^{\dagger}_{j} f_{i} 
+
\underbrace{k \rightarrow j \rightarrow i}
+ 
\underbrace{j \rightarrow i \rightarrow k} \; + \;  \rm{h.c.} 
\end{align}
Note that in the last line, we removed the spin-label. Since the spinon bands are degenerate in the spin degrees of freedom, we have added the contributions from both spin channels. In a similar spirit, we can re-express the charge fluctuation operator in spinon language as~\cite{Paper11}
\begin{equation}\label{Eq.20}
\delta \rho_{i,jk} = e\frac{8t^3}{U^3}  \left( \mathbf{S}_i \cdot \mathbf{S}_j + \mathbf{S}_i \cdot \mathbf{S}_k - 2 \mathbf{S}_j \cdot \mathbf{S}_k \right) 
				   = e\frac{8m_0 t^3}{U^3}  \left( e^{i\phi_{ji}}f^{\dagger}_{i} f_{j} 
																		+    e^{i\phi_{ki}}f^{\dagger}_{i} f_{k} 
																		-  2 e^{i\phi_{kj}}f^{\dagger}_{j} f_{k} \right) + \rm{h.c.},
\end{equation}
where we again added the spin degeneracy in the last line of the above equation.

\begin{figure}[t]
\centering
\includegraphics[width=1\linewidth]{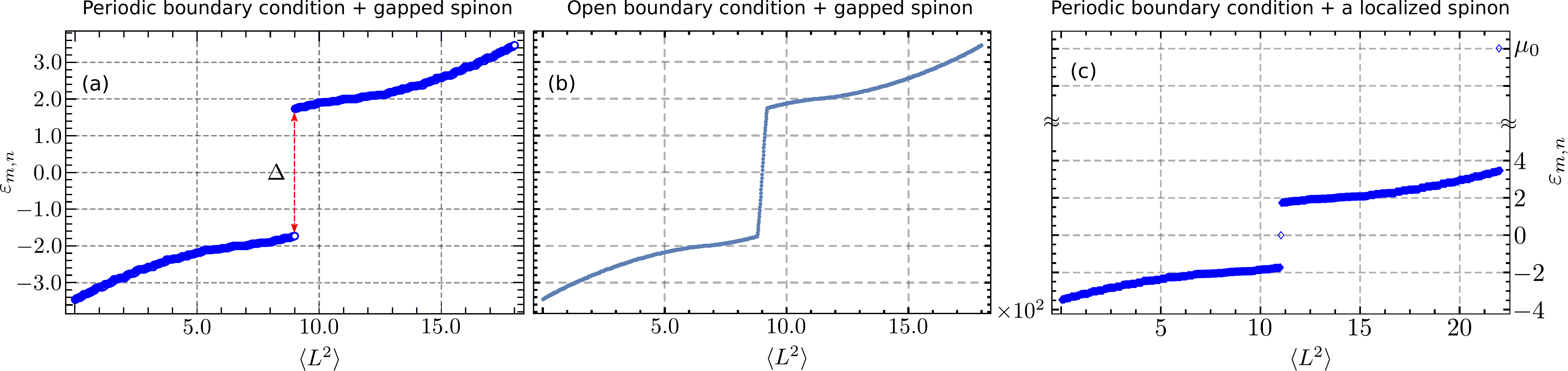} 
\caption{(a) The eigenvalue distribution of the real-space Hamiltonian (PBC) as a function of the system size. The topological protection manifests into a gap in the eigenvalue distribution. (b) The eigenvalue distribution of the real-space Hamiltonian as a function of the system size in open boundary conditions with the edge modes in the gap. (c) The eigenvalue distribution of the real-space Hamiltonian as a function of the system size for the triangular lattice in periodic boundary conditions. We set a local chemical potential $V_0$ at site $\dr_0$ for the A-sublattice. For very large $V_0$, there are two additional states which are shown by the symbols around zero energy and the other one at higher energy dictated by $V_0$.}\label{fig:SFig2}
\end{figure} 

As we are interested in the real space loop current and chare fluctuation profile, we consider a real-space calculation to evaluate the loop current expectation values. The Hamiltonian in Eq.~\eqref{Eq.15} is therefore written on a 2D triangle lattice of size $L \times L$ unit cells. Since each unit cell contains two sub-lattice sites (A \& B), there are $2L^2$ spinon operators $f^{\dagger}_{i\sigma}$ in the system. The lattice vectors are chosen as $\mathbf{a}_1 = (2, 0)$, and $\mathbf{a}_2 = (1/2,\sqrt{3}/2)$. The spinon operators at a site $i$ is written as $f^{\dagger}_i = f^{\dagger}_{\eta} (m,n)$, where $\eta$ corresponds to the sub-lattice index, and $\mathbf{R}(m,n) = m \mathbf{a}_1 + n \mathbf{a}_2$, $m,n = 1,2,\ldots L$. If we impose periodic boundary condition (PBC), then the spinon operators follow $f^{\dagger}_{\eta}(m+L,n) = f^{\dagger}_{\eta}(m,n)$, and $f^{\dagger}_{\eta}(m,n+L) = f^{\dagger}_{\eta}(m,n)$, otherwise $f^{\dagger}_{\eta}(m+L,n) = f^{\dagger}_{\eta}(m,n+L) = 0$ in the open boundary condition (OBC). The $2L^2$-dimensional spinon vector is constructed as $\mathbf{\tilde{f}^{\dagger}} = (f^{\dagger}_{\mathrm{A}}, f^{\dagger}_{\mathrm{B}})$ with
\begin{equation}\label{Eq.21}
f^{\dagger}_{\eta} = \left( f^{\dagger}_\eta(1,1), f^{\dagger}_\eta(2,1), \ldots f^{\dagger}_\eta(L,1), f^{\dagger}_\eta(1,2), f^{\dagger}_\eta(2,2), \ldots f^{\dagger}_\eta(L,2), \ldots, f^{\dagger}_\eta(L,L) \right), \quad \eta \in \{ \mathrm{A}, \mathrm{B} \}.
\end{equation}
In terms of the spinon vector $\mathbf{\tilde{f}^{\dagger}}$, we can write the the Hamiltonian in Eq.~\eqref{Eq.15} as $\mathcal{H}_{\rm{eff}} = \mathbf{\tilde{f}}^{\dagger} \mathsf{H} \mathbf{\tilde{f}}$, where $\mathsf{H}$ is written as a $2L^2 \times 2L^2$ matrix. On the diagonal basis, we can rewrite the spinon operators as
\begin{equation}\label{Eq.22}
\begin{bmatrix}
\zeta_{u\sigma} (m,n) \\
\zeta_{d\sigma} (m,n) 
\end{bmatrix} = 
\sum_{m,n} \mathrm{U}_{mn} \cdot
\begin{bmatrix}
f_{\mathrm{A} \sigma} (m,n) \\
f_{\mathrm{B} \sigma} (m,n) 
\end{bmatrix},
\end{equation}
where $\mathrm{U}$ is the diagonalizing matrix for Hamiltonian $\sf{H}$, and $\zeta_{u/d}(m,n)$ are the diagonal spinon operators corresponding to the spectrum as shown in Fig.~\ref{fig:SFig2}. In panel (a), we show the band dispersion with a topological gap $\Delta$ in the PBC, while in panel (b), the spectrum in the case of OBC is shown with an edge mode inside the bulk gap $\Delta$. The spectrum for a localized spinon at site $\dr_0$ inside the bulk is shown in panel (c). To realize the latter scenario, we impose a large onsite chemical potential $V_0$ at the site $\dr_0$ within the unit cell. Due to the large energy, this specific site will not host any spinons and can be thought of as a localized spinon hole. The physical situation might be some empty defect sites, or some magnetic impurity sitting inside the bulk of the system. \\

{\bf{Analysis of the expectation values}} \\

Once we know all the eigenenergy and the eigenstates of the Hamiltonian, it is straightforward to obtain the average of the loop current operator in the ground state. The target quantity of our interest is $\braket{f^{\dagger}_\alpha(\mathbf{R}_{mn}) f_\beta(\mathbf{R}'_{m'n'})}$ for $\{ \alpha, \beta \} \in \{ \mathrm{A}, \mathrm{B} \}$, where $\mathbf{R}_{mn}$ denotes the position of the site at $\mathbf{R}_{mn} = m\hat{\mathbf{a}}_1 + n \hat{\mathbf{a}}_2$. The analysis goes as follows
\begin{equation}\label{Eq.23}
\braket{f^{\dagger}_{\alpha}(\mathbf{R}_{mn}) f_{\beta}(\mathbf{R}'_{m'n'})}
=
\big \langle \begin{pmatrix}
f^{\dagger}_{\mathrm{A}} & f^{\dagger}_{\mathrm{B}} 
\end{pmatrix}
\mathcal{W}^{(\alpha \beta)}_{\mathbf{R} \mathbf{R}'}
\begin{pmatrix}
f_{\mathrm{A}} \\
f_{\mathrm{B}} 
\end{pmatrix}
\big \rangle
=
\sum_{ll'=1}^{2L^2} 
\left( U^{\dagger} \mathcal{W}^{(\alpha \beta)}_{\mathbf{R} \mathbf{R}'} U \right)_{ll'} \braket{\zeta^{\dagger}_l \zeta_{l'}}
=
\sum_{l=L^2+1}^{2L^2} 
\left( U^{\dagger} \mathcal{W}^{(\alpha \beta)}_{\mathbf{R} \mathbf{R}'} U \right)_{ll},
\end{equation}
where $\braket{\zeta^{\dagger}_l \zeta_{l'}} = \delta_{ll'}$, and the $2L^2 \times 2L^2$ matrices $\mathcal{W}^{(\alpha \beta)}_{\mathbf{R} \mathbf{R}'}$ are defined as follows
\begin{equation}\label{Eq.24}
\mathcal{W}^{(\alpha \beta)}_{\mathbf{R} \mathbf{R}'}
=
\begin{pmatrix}
\renewcommand*{\arraystretch}{3}
\mathcal{B}^{\mathrm{A A}}_{\mathbf{R} \mathbf{R}'}	\delta_{\alpha \mathrm{A}} \delta_{\beta \mathrm{A}} 		&		\mathcal{B}^{\mathrm{A B}}_{\mathbf{R} \mathbf{R}'}	\delta_{\alpha \mathrm{A}} \delta_{\beta \mathrm{B}}		\\
\mathcal{B}^{\mathrm{B A}}_{\mathbf{R} \mathbf{R}'}	\delta_{\alpha \mathrm{B}} \delta_{\beta \mathrm{A}}		&	 	\mathcal{B}^{\mathrm{BB}}_{\mathbf{R} \mathbf{R}'} \delta_{\alpha \mathrm{B}} \delta_{\beta \mathrm{B}}
\end{pmatrix}.
\end{equation}
Here, $\mathcal{B}^{(\alpha)}_{\mathbf{R} \mathbf{R}'}$ are $L^2 \times L^2$ matrices corresponding to the non-zero connections allowed by the orientations of the triangles. The corresponding results for the charge fluctuation and loop current distribution are shown in Fig.~3(a-c) in the main text.

\section{Dynamic spin structure factor \label{sec:sec.5}}

In this section, we provide the details of the analysis of the dynamical spin structure factor (DSSF). The latter is defined as
\begin{equation}\label{Eq.25}
S(\vq,\omega) 
=
\sum_{i,j} \frac{e^{i\vq \cdot(\dr_i - \dr_j)}}{N_{s}} \int_{-\infty}^{\infty} dt e^{i\omega t } \braket{\mathbf{S}_{i}(t) \cdot \mathbf{S}_j(0)},
\end{equation}
where $i,j$ corresponds to the position of the unit cell containing two sub-lattice sites. Note that the unit-cell $i$ has two sub-lattice sites labeled by A, and B. Consequently, we can rewrite the above equation as
\begin{align}
\nonumber
S(\vq,\omega) 
=
\sum_{i,j} \frac{e^{i\vq \cdot(\dr_i - \dr_j)}}{N_{s}} \int_{-\infty}^{\infty} dt e^{i\omega t } 
& \Big[ \braket{\mathbf{S}_{i,{\rm{A}}}(t) \cdot \mathbf{S}_{j,{\rm{A}}}(0)} + \braket{\mathbf{S}_{i,{\rm{B}}}(t) \cdot \mathbf{S}_{j,{\rm{B}}}(0)} + \\
\label{Eq.26}
&
e^{i\vq \cdot {\bf{a}_1}}\braket{\mathbf{S}_{i,{\rm{A}}}(t) \cdot \mathbf{S}_{j,{\rm{B}}}(0)}
+ e^{-i\vq \cdot {\bf{a}_1}}\braket{\mathbf{S}_{i,{\rm{B}}}(t) \cdot \mathbf{S}_{j,{\rm{A}}}(0)} \Big], 
\end{align}
where we have explicitly written down the DSSF in sub-lattice resolved coordinates. It is straightforward to show next that a typical sub-lattice resolved term is given by ($\{ \eta, \zeta \} \in \{ \rm{A,B} \}$)
\begin{align}
\nonumber
S^{\eta \zeta}(\vq,\omega) 
=
&
\sum_{i,j} \frac{e^{i\vq \cdot(\dr_i - \dr_j)}}{4N_{s}} \int_{-\infty}^{\infty} dt e^{i\omega t} e^{i (E_0 - E_n)t} \braket{0|f^{\dag}_{i \eta \alpha} f_{i \eta \beta}|n}\braket{n|f^{\dag}_{j \zeta \gamma} f_{j \zeta \delta}|0} (2 \delta_{\alpha \delta} \delta_{\beta \gamma} - \delta_{\alpha \beta} \delta_{\gamma \delta}) \\
\nonumber
&
= \sum_{\vk, \vp} \sum_n \frac{2 \delta_{\alpha \delta} \delta_{\beta \gamma} - \delta_{\alpha \beta} \delta_{\gamma \delta}}{4} \delta(\omega - E_n + E_0)
\braket{0|f^{\dag}_{\vp \eta \alpha} f_{ \vp + \vq \eta \beta}|n}\braket{n|f^{\dag}_{ \vk + \vq \zeta \gamma} f_{ \vk \zeta \delta}|0} \\
\nonumber
&
= \sum_{\vk} \sum_n  \frac{2 \delta_{\alpha \delta} \delta_{\beta \gamma} - \delta_{\alpha \beta} \delta_{\gamma \delta}}{4} \delta(\omega - E_n + E_0)
\braket{0|f^{\dag}_{\vk \eta \alpha} f_{ \vk + \vq \eta \beta}|n}\braket{n|f^{\dag}_{ \vk + \vq \zeta \gamma} f_{ \vk \zeta \delta}|0} \\
\nonumber
&
= \sum_{\vk} \sum_n  \delta(\omega - E_n + E_0)
\left( \frac{\braket{0|f^{\dag}_{\vk \eta \alpha} f_{ \vk + \vq \eta \beta}|n}\braket{n|f^{\dag}_{ \vk + \vq \zeta \beta} f_{ \vk \zeta \alpha}|0}}{2} - \frac{\braket{0|f^{\dag}_{\vk \eta \alpha} f_{ \vk + \vq \eta \alpha}|n}\braket{n|f^{\dag}_{ \vk + \vq \zeta  \beta} f_{ \vk \zeta \beta}|0}}{4} \right) \\
\nonumber
&
=
\sum_{\vk} \sum_n  \delta(\omega - E_n + E_0)
\left( \frac{\braket{0|f^{\dag}_{\vk \eta \alpha} f_{ \vk + \vq \eta \beta}|n}\braket{n|f^{\dag}_{ \vk + \vq \zeta \beta} f_{ \vk \zeta \alpha}|0}}{2} - \frac{\braket{0|f^{\dag}_{\vk \eta \alpha} f_{ \vk + \vq \eta \alpha}|n}\braket{n|f^{\dag}_{ \vk + \vq \zeta \alpha} f_{ \vk \zeta \alpha}|0}}{4} \right) \\
\label{Eq.27}
&
=
\frac{3}{2} \sum_{\vk} \sum_n  \delta(\omega - E_n + E_0)
\braket{0|f^{\dag}_{\vk \eta} f_{ \vk + \vq \eta}|n}\braket{n|f^{\dag}_{ \vk + \vq \zeta} f_{ \vk \zeta}|0},
\end{align}
where in the last line, we have summed over the degenerate spin degrees of freedom and omitted the spin indices, and $\ket{n}$ corresponds to an excited eigenmode with energy $E_n$. We perform the numerical integration in Mathematica and the corresponding plots are shown in Fig.~4 in the main text. We approximate the delta function $\delta(x)$ as $\delta(x) = \tfrac{1}{\pi} \tfrac{\gamma^2}{x^2 + \gamma^2}$, and considered $\gamma = 0.1$ for numerical purposes.

\section{Optical conductivity in the CSL phase \label{sec:sec.6}}

\begin{figure}[t]
\centering
\includegraphics[width=0.2\linewidth]{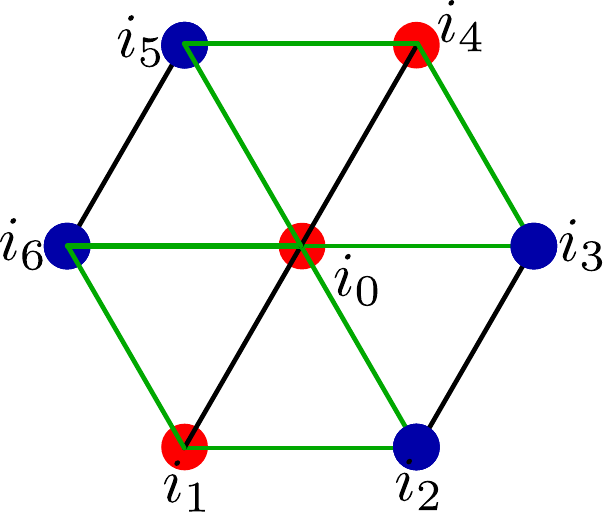} 
\caption{Contribution to the polarization from a single site at $i_0$ surrounded by six triangles with sites $i_1, \ldots, i_6$.}\label{fig:SFig3}
\end{figure}

Finally, in this section, we provide the details of the analysis for the transverse and longitudinal optical conductivity in the CSL phase. Since the parent compound is a Mott insulator, we do not have any mobile charges; however, the charge fluctuations in the insulating phase will lead to electrical polarization which couples to the external electric field and lead to finite optical conductivity. The corresponding electrical susceptibility is given by
\begin{equation}\label{Eq.28}
\chi_{xy} (\omega) 
= -
\frac{V}{\hbar} \sum_{n \neq 0} 
\Bigg[ \frac{\braket{\psi_0| P_{x}| \psi_n} \braket{\psi_n | P_{y}| \psi_0}}{\omega - \omega_n + i\eta} - 
\frac{\braket{\psi_0| P_{y}| \psi_n} \braket{\psi_n | P_{x}| \psi_0}}{\omega + \omega_n + i\eta} \Bigg],
\end{equation}
where $\mathbf{P} = P_x \hat{\bf{x}} + P_y \hat{\bf{y}}$ is the total polarization in the system. For a triangular plaquette, the corresponding expression can be obtained from the charge fluctuation operators~\cite{Paper11}. Here, we consider a single site at $i_0$ embedded in the lattice as shown in Fig.~\ref{fig:SFig3}. The polarization for each triangle is thereafter written as 
\begin{subequations}
\begin{align}
\label{Eq.29.1}
\mathbf{P}_{i_0;i_1i_2} & 
= 
- \frac{12 eat^3}{U^3} \left( \vs_{i_0} \cdot \vs_{i_1} - \vs_{i_0} \cdot \vs_{i_2} \right) \mathbf{\hat{x}} 
+ \frac{4\sqrt{3} eat_{{\rm{hop}}}^3}{U^3} \left( \vs_{i_0} \cdot \vs_{i_1} + \vs_{i_0} \cdot \vs_{i_2} - 2 \vs_{i_1} \cdot \vs_{i_2} \right) \mathbf{\hat{y}}, \\
\label{Eq.29.2}
\mathbf{P}_{i_0;i_4i_5} & 
= 
 \frac{12 eat^3}{U^3} \left( \vs_{i_0} \cdot \vs_{i_4} - \vs_{i_0} \cdot \vs_{i_5} \right) \mathbf{\hat{x}} 
- \frac{4\sqrt{3} eat_{{\rm{hop}}}^3}{U^3} \left( \vs_{i_0} \cdot \vs_{i_4} + \vs_{i_0} \cdot \vs_{i_5} - 2 \vs_{i_4} \cdot \vs_{i_5} \right) \mathbf{\hat{y}}, \\
\label{Eq.29.3}
\mathbf{P}_{i_0;i_2i_3} & 
= 
 \frac{12 eat^3}{U^3} \left( \vs_{i_2} \cdot \vs_{i_3} - \vs_{i_0} \cdot \vs_{i_2} \right) \mathbf{\hat{x}} 
- \frac{4\sqrt{3} eat_{{\rm{hop}}}^3}{U^3} \left( \vs_{i_0} \cdot \vs_{i_2} + \vs_{i_2} \cdot \vs_{i_3} - 2 \vs_{i_0} \cdot \vs_{i_3} \right) \mathbf{\hat{y}}, \\
\label{Eq.29.4}
\mathbf{P}_{i_0;i_5i_6} & 
= 
- \frac{12 eat^3}{U^3} \left( \vs_{i_5} \cdot \vs_{i_6} - \vs_{i_0} \cdot \vs_{i_5} \right) \mathbf{\hat{x}} 
+ \frac{4\sqrt{3} eat_{{\rm{hop}}}^3}{U^3} \left( \vs_{i_0} \cdot \vs_{i_5} + \vs_{i_5} \cdot \vs_{i_6} - 2 \vs_{i_0} \cdot \vs_{i_6} \right) \mathbf{\hat{y}}, \\
\label{Eq.29.5}
\mathbf{P}_{i_0;i_3i_4} & 
= 
 \frac{12 eat^3}{U^3} \left( \vs_{i_3} \cdot \vs_{i_4} - \vs_{i_0} \cdot \vs_{i_4} \right) \mathbf{\hat{x}} 
+ \frac{4\sqrt{3} eat_{{\rm{hop}}}^3}{U^3} \left( \vs_{i_0} \cdot \vs_{i_4} + \vs_{i_3} \cdot \vs_{i_4} - 2 \vs_{i_0} \cdot \vs_{i_3} \right) \mathbf{\hat{y}}, \\
\label{Eq.29.6}
\mathbf{P}_{i_0;i_6i_1} & 
= 
- \frac{12 eat^3}{U^3} \left( \vs_{i_6} \cdot \vs_{i_1} - \vs_{i_0} \cdot \vs_{i_1} \right) \mathbf{\hat{x}} 
- \frac{4\sqrt{3} eat_{{\rm{hop}}}^3}{U^3} \left( \vs_{i_0} \cdot \vs_{i_1} + \vs_{i_6} \cdot \vs_{i_1} - 2 \vs_{i_0} \cdot \vs_{i_6} \right) \mathbf{\hat{y}}.
\end{align}
\end{subequations}
Now, we can add all these contributions to obtain the total polarization per site $i_0$ as 
\begin{align}
\nonumber
\mathbf{P}_{i_0}
& 
=
\frac{12 eat^3}{U^3} \left( \vs_{i_2} \cdot \vs_{i_3} - \vs_{i_5} \cdot \vs_{i_6} + \vs_{i_3} \cdot \vs_{i_4} - \vs_{i_6} \cdot \vs_{i_1}\right) \mathbf{\hat{x}} \\
\label{Eq.30}
& - \frac{4\sqrt{3} eat^3}{U^3} \left( 2 \vs_{i_1} \cdot \vs_{i_2} + \vs_{i_2} \cdot \vs_{i_3} - \vs_{i_3} \cdot \vs_{i_4} - 2 \vs_{i_4} \cdot \vs_{i_5} - \vs_{i_5} \cdot \vs_{i_6} + \vs_{i_6} \cdot \vs_{i_1} \right) \mathbf{\hat{y}}.
\end{align}
We further utlize the Pauli matrix identities and rewrite the above expression in the spinon language as 
\begin{align}
\nonumber
\mathbf{P}_{i_0}
=
\frac{3 eat^3}{U^3} & \delta_{\alpha \beta \gamma \delta} \left( 
							f^{\dag}_{i_2\alpha} f^{\dag}_{i_3\gamma} f_{i_3\delta} f_{i_2\beta} 
							- f^{\dag}_{i_5\alpha} f^{\dag}_{i_6\gamma} f_{i_6\delta} f_{i_5\beta}
							+ f^{\dag}_{i_3\alpha} f^{\dag}_{i_4\gamma} f_{i_4\delta} f_{i_3\beta}
							- f^{\dag}_{i_6\alpha} f^{\dag}_{i_1\gamma} f_{i_1\delta} f_{i_6\beta} \right) \hat{\bf{x}} - 
							\frac{\sqrt{3} eat^3}{U^3} \delta_{\alpha \beta \gamma \delta} \left( 2 f^{\dag}_{i_1\alpha} f^{\dag}_{i_2\gamma} f_{i_2\delta} f_{i_1\beta} \right.\\
\label{Eq.31}
& 							\left.
						    + f^{\dag}_{i_2\alpha} f^{\dag}_{i_3\gamma} f_{i_3\delta} f_{i_2\beta}
							- f^{\dag}_{i_3\alpha} f^{\dag}_{i_4\gamma} f_{i_4\delta} f_{i_3\beta}
							- 2 f^{\dag}_{i_4\alpha} f^{\dag}_{i_5\gamma} f_{i_5\delta} f_{i_4\beta}
							- f^{\dag}_{i_5\alpha} f^{\dag}_{i_6\gamma} f_{i_6\delta} f_{i_5\beta}
							+ f^{\dag}_{i_6\alpha} f^{\dag}_{i_1\gamma} f_{i_1\delta} f_{i_6\beta}	\right) \hat{\bf{y}},
\end{align}
where $\delta_{\alpha \beta \gamma \delta} = 2 \delta_{\alpha \delta} \delta_{\beta \gamma} - \delta_{\alpha \beta} \delta_{\gamma \delta}$. Consequently, we can obtain the total polarization which is the sum over two sub-lattice polarizations as
\begin{equation}\label{Eq.32}
\mathbf{P} = \frac{1}{V} \left( \sum_{i_0 \in {\rm{A}}} \mathbf{P}_{i_0} + \sum_{i_0 \in {\rm{B}}} \mathbf{P}_{i_0} \right).
\end{equation}
Rewriting it in the momentum space we obtain in the sub-lattice basis as 
\begin{subequations}
\begin{align}
\nonumber
P^x_{i_0 \in {\rm{A}}} 
& 
= 
\frac{3 eat^3}{U^3} \sum_{\{ \vk \}} \delta_{\alpha \beta \gamma \delta} 
				 \left( e^{i \vk_1 \cdot \dr_2 + i \vk_2 \cdot \dr_3 - i \vk_3 \cdot \dr_3 - i \vk_4 \cdot \dr_2} f^{\dag}_{{\rm{B}}\vk_1\alpha} f^{\dag}_{{\rm{B}}\vk_2\gamma} f_{{\rm{B}}\vk_3\delta} f_{{\rm{B}}\vk_4\beta} 
					  - e^{i \vk_1 \cdot \dr_5 + i \vk_2 \cdot \dr_6 - i \vk_3 \cdot \dr_6 - i \vk_4 \cdot \dr_5} f^{\dag}_{{\rm{B}}\vk_1\alpha} f^{\dag}_{{\rm{B}}\vk_2\gamma} f_{{\rm{B}}\vk_3\delta} f_{{\rm{B}}\vk_4\beta} \right. \\
\label{Eq.33.1}
&
\left. \qquad         + e^{i \vk_1 \cdot \dr_3 + i \vk_2 \cdot \dr_4 - i \vk_3 \cdot \dr_4 - i \vk_4 \cdot \dr_3} f^{\dag}_{{\rm{B}}\vk_1\alpha} f^{\dag}_{{\rm{A}}\vk_2\gamma} f_{{\rm{A}}\vk_3\delta} f_{{\rm{B}}\vk_4\beta}
					  - e^{i \vk_1 \cdot \dr_6 + i \vk_2 \cdot \dr_1 - i \vk_3 \cdot \dr_1 - i \vk_4 \cdot \dr_6} f^{\dag}_{{\rm{B}}\vk_1\alpha} f^{\dag}_{{\rm{A}}\vk_2\gamma} f_{{\rm{A}}\vk_3\delta} f_{{\rm{B}}\vk_4\beta} \right), \\
\nonumber
P^y_{i_0 \in {\rm{A}}} 
& 
= 
\frac{\sqrt{3} eat^3}{U^3} \sum_{\{ \vk \}} \delta_{\alpha \beta \gamma \delta}
				 \left( 2 e^{i \vk_1 \cdot \dr_1 + i \vk_2 \cdot \dr_2 - i \vk_3 \cdot \dr_2 - i \vk_4 \cdot \dr_1} f^{\dag}_{{\rm{A}}\vk_1\alpha} f^{\dag}_{{\rm{B}}\vk_2\gamma} f_{{\rm{B}}\vk_3\delta} f_{{\rm{A}}\vk_4\beta} 
					  + e^{i \vk_1 \cdot \dr_2 + i \vk_2 \cdot \dr_3 - i \vk_3 \cdot \dr_3 - i \vk_4 \cdot \dr_2} f^{\dag}_{{\rm{B}}\vk_1\alpha} f^{\dag}_{{\rm{B}}\vk_2\gamma} f_{{\rm{B}}\vk_3\delta} f_{{\rm{B}}\vk_4\beta} \right. \\		
\nonumber
&
\left. \qquad		  - e^{i \vk_1 \cdot \dr_3 + i \vk_2 \cdot \dr_4 - i \vk_3 \cdot \dr_4 - i \vk_4 \cdot \dr_3} f^{\dag}_{{\rm{B}}\vk_1\alpha} f^{\dag}_{{\rm{A}}\vk_2\gamma} f_{{\rm{A}}\vk_3\alpha} f_{{\rm{B}}\vk_4\beta}
					  - 2 e^{i \vk_1 \cdot \dr_4 + i \vk_2 \cdot \dr_5 - i \vk_3 \cdot \dr_5 - i \vk_4 \cdot \dr_4} f^{\dag}_{{\rm{A}}\vk_1\alpha} f^{\dag}_{{\rm{B}}\vk_2\gamma} f_{{\rm{B}}\vk_3\alpha} f_{{\rm{A}}\vk_4\beta} \right. \\					    
\label{Eq.33.2}
&
\left. \qquad         -	e^{i \vk_1 \cdot \dr_5 + i \vk_2 \cdot \dr_6 - i \vk_3 \cdot \dr_6 - i \vk_4 \cdot \dr_5} f^{\dag}_{{\rm{B}}\vk_1\alpha} f^{\dag}_{{\rm{B}}\vk_2\gamma} f_{{\rm{B}}\vk_3\delta} f_{{\rm{B}}\vk_4\beta}
					  + e^{i \vk_1 \cdot \dr_6 + i \vk_2 \cdot \dr_1 - i \vk_3 \cdot \dr_1 - i \vk_4 \cdot \dr_6} f^{\dag}_{{\rm{B}}\vk_1\alpha} f^{\dag}_{{\rm{A}}\vk_2\gamma} f_{{\rm{A}}\vk_3\delta} f_{{\rm{B}}\vk_4\beta} \right),
	\end{align}
\end{subequations}
whereas for $i_0 \in \rm{B}$, we need to update the above expression with $\rm{A} \leftrightarrow \rm{B}$. Now, we utilize the vector relation as $\dr_i = \dr_{i_0} + \bm{\Delta}_i, \forall i = 1, \ldots 6$, where $\bm{\Delta}_i$ is the nearest neighbor site to $i_0$, and they are related to the original nearest-neighbor vectors defined as
\begin{equation}\label{Eq.34}
\bm{\Delta}_1 = - \bm{\delta}_2, \quad
\bm{\Delta}_2 =   \bm{\delta}_3, \quad
\bm{\Delta}_3 =   \bm{\delta}_1, \quad
\bm{\Delta}_4 =   \bm{\delta}_2, \quad
\bm{\Delta}_5 = - \bm{\delta}_3, \quad 
\bm{\Delta}_6 = - \bm{\delta}_1. 
\end{equation}
Taking the summation over all the sites $i_0$, we have
\begin{subequations}
\begin{align}
\nonumber
\sum_{i_0} P^x_{i_0 \in {\rm{A}}}  
&
= 
\frac{6 ieat^3}{U^3} \sum'_{\{ \vk \}} \delta_{\alpha \beta \gamma \delta} 
 \left( \sin{[\vk_1 \cdot \bm{\delta}_3 + \vk_2 \cdot \bm{\delta}_1 - \vk_3 \cdot \bm{\delta}_1 - \vk_4 \cdot \bm{\delta}_3]} f^{\dag}_{{\rm{B}}\vk_1\alpha}f^{\dag}_{{\rm{B}}\vk_2\gamma} f_{{\rm{B}}\vk_3\delta} f_{{\rm{B}}\vk_4\beta} \;+ \right. \\
\label{Eq.35.1}
&
\left. \qquad \qquad \qquad \qquad \quad 	
	   \sin{[\vk_1 \cdot \bm{\delta}_1 + \vk_2 \cdot \bm{\delta}_2 - \vk_3 \cdot \bm{\delta}_2 - \vk_4 \cdot \bm{\delta}_1]} f^{\dag}_{{\rm{B}}\vk_1\alpha} f^{\dag}_{{\rm{A}}\vk_2\gamma} f_{{\rm{A}}\vk_3\delta} f_{{\rm{B}}\vk_4\beta}  \right), \\	  
\nonumber
\sum_{i_0} P^y_{i_0 \in {\rm{A}}} 
&
= 
\frac{2\sqrt{3} ieat^3}{U^3} \sum'_{\{ \vk \}} \delta_{\alpha \beta \gamma \delta} 
\left(2 \sin[-\vk_1 \cdot \bm{\delta}_2 + \vk_2 \cdot \bm{\delta}_3 - \vk_3 \cdot \bm{\delta}_3 + \vk_4 \cdot \bm{\delta}_2] f^{\dag}_{{\rm{A}}\vk_1\alpha} f^{\dag}_{{\rm{B}}\vk_2\gamma} f_{{\rm{B}}\vk_3\delta} f_{{\rm{A}}\vk_4\beta} \;+ \right. \\
\nonumber
&
\left.  \qquad \qquad \qquad \qquad \qquad
\sin[\vk_1 \cdot \bm{\delta}_3 + \vk_2 \cdot \bm{\delta}_1 - \vk_3 \cdot \bm{\delta}_1 - \vk_4 \cdot \bm{\delta}_3] f^{\dag}_{{\rm{B}}\vk_1\alpha} f^{\dag}_{{\rm{B}}\vk_2\gamma} f_{{\rm{B}}\vk_3\delta} f_{{\rm{B}}\vk_4\beta} \;-  \right. \\
\label{Eq.35.2}
&
\left.  \qquad \qquad \qquad \qquad \qquad \quad
\sin[\vk_1 \cdot \bm{\delta}_1 + \vk_2 \cdot \bm{\delta}_2 - \vk_3 \cdot \bm{\delta}_2 - \vk_4 \cdot \bm{\delta}_1] f^{\dag}_{{\rm{B}}\vk_1\alpha} f^{\dag}_{{\rm{A}}\vk_2\gamma} f_{{\rm{A}}\vk_3\delta} f_{{\rm{B}}\vk_4\beta} \right),
	\end{align}
\end{subequations}
where $\sum'$ denotes the momentum conversation, \textit{i.e.}, $\vk_1 + \vk_2 = \vk_3 + \vk_4$, obtained by summing over all the sites.
\begin{subequations}
\begin{align}
\nonumber
P^x & =\frac{6 ieat^3}{U^3} \sum'_{\{ \vk \}} \delta_{\alpha \beta \gamma \delta}
 \left( \sin{[\vk_1 \cdot \bm{\delta}_3 + \vk_2 \cdot \bm{\delta}_1 - \vk_3 \cdot \bm{\delta}_1 - \vk_4 \cdot \bm{\delta}_3]} 
 							( f^{\dag}_{{\rm{B}}\vk_1\alpha}f^{\dag}_{{\rm{B}}\vk_2\gamma} f_{{\rm{B}}\vk_3\delta} f_{{\rm{B}}\vk_4\beta} +
 							f^{\dag}_{{\rm{A}}\vk_1\alpha}f^{\dag}_{{\rm{A}}\vk_2\gamma} f_{{\rm{A}}\vk_3\delta} f_{{\rm{A}}\vk_4\beta})  \;+ \right. \\
\label{Eq.36.1}
&
\left. \qquad \qquad \qquad \qquad 
	    \sin{[\vk_1 \cdot \bm{\delta}_1 + \vk_2 \cdot \bm{\delta}_2 - \vk_3 \cdot \bm{\delta}_2 - \vk_4 \cdot \bm{\delta}_1]} 
					    (f^{\dag}_{{\rm{B}}\vk_1\alpha} f^{\dag}_{{\rm{A}}\vk_2\gamma} f_{{\rm{A}}\vk_3\delta} f_{{\rm{B}}\vk_4\beta} + 
   					     f^{\dag}_{{\rm{A}}\vk_1\alpha} f^{\dag}_{{\rm{B}}\vk_2\gamma} f_{{\rm{B}}\vk_3\delta} f_{{\rm{A}}\vk_4\beta}) \right), \\
\nonumber	
P^y & = \frac{2\sqrt{3} ieat^3}{U^3} \sum'_{\{ \vk \}} \delta_{\alpha \beta \gamma \delta}
 \left( 2 \sin[-\vk_1 \cdot \bm{\delta}_2 + \vk_2 \cdot \bm{\delta}_3 - \vk_3 \cdot \bm{\delta}_3 + \vk_4 \cdot \bm{\delta}_2] 
 \left( f^{\dag}_{{\rm{A}}\vk_1\alpha} f^{\dag}_{{\rm{B}}\vk_2\gamma} f_{{\rm{B}}\vk_3\delta} f_{{\rm{A}}\vk_4\beta} + f^{\dag}_{{\rm{B}}\vk_1\alpha} f^{\dag}_{{\rm{A}}\vk_2\gamma} f_{{\rm{A}}\vk_3\delta} f_{{\rm{B}}\vk_4\beta} \right) \; + \right. \\
\nonumber
&
 \left. \qquad \qquad
  \sin[\vk_1 \cdot \bm{\delta}_3 + \vk_2 \cdot \bm{\delta}_1 - \vk_3 \cdot \bm{\delta}_1 - \vk_4 \cdot \bm{\delta}_3] 
  \left( f^{\dag}_{{\rm{B}}\vk_1\alpha} f^{\dag}_{{\rm{B}}\vk_2\gamma} f_{{\rm{B}}\vk_3\delta} f_{{\rm{B}}\vk_4\beta} + f^{\dag}_{{\rm{A}}\vk_1\alpha} f^{\dag}_{{\rm{A}}\vk_2\gamma} f_{{\rm{A}}\vk_3\delta} f_{{\rm{A}}\vk_4\beta} \right) \; - \right. \\
\label{Eq.36.2}
&
 \left. \qquad \qquad \qquad  
  \sin[\vk_1 \cdot \bm{\delta}_1 + \vk_2 \cdot \bm{\delta}_2 - \vk_3 \cdot \bm{\delta}_2 - \vk_4 \cdot \bm{\delta}_1] 
  \left( f^{\dag}_{{\rm{B}}\vk_1\alpha} f^{\dag}_{{\rm{A}}\vk_2\gamma} f_{{\rm{A}}\vk_3\delta} f_{{\rm{B}}\vk_4\beta} + f^{\dag}_{{\rm{A}}\vk_1\alpha} f^{\dag}_{{\rm{B}}\vk_2\gamma} f_{{\rm{B}}\vk_3\delta} f_{{\rm{A}}\vk_4\beta} \right) \right).
\end{align}
\end{subequations}

%

\end{document}